\documentclass[prd,notitlepage,twocolumn,nofootinbib,superscriptaddress,altaffilletter, preprintnumbers,floatfix]{revtex4-1}
\pdfoutput=1
\usepackage[english]{babel}
\usepackage{amsmath,amssymb,amsfonts, bm,slashed, subdepth, bbold}
\usepackage{graphicx}
\usepackage[sort&compress]{natbib}
\usepackage{xcolor}
\usepackage[normalem]{ulem}
\usepackage{hyperref}
\usepackage{cleveref}
\definecolor{red}{rgb}{1.0, 0, 0}
\usepackage{hyperref}
\usepackage{enumerate}
\usepackage{epsfig, subfigure}
\usepackage{setspace}
\usepackage{booktabs, tabularx}
\usepackage{units}
\usepackage{hyperref}
\usepackage[utf8]{inputenc}

\newcommand{\be}{\begin{equation}}
\newcommand{\ee}{\end{equation}}
\newcommand{\ba}{\begin{array}}
\newcommand{\ea}{\end{array}}
\newcommand{\bea}{\begin{eqnarray}}
\newcommand{\eea}{\end{eqnarray}}
\newcommand{\balg}{\begin{align}}
\newcommand{\ealg}{\end{align}}
\newcommand{\bit}{\begin{itemize}}
\newcommand{\eit}{\end{itemize}}
\newcommand{\trm}[1]{\textrm{#1}}

\newcommand{\Mpc}{\trm{\Mpc}}
\newcommand{\yr}{\trm{\yr}}
\newcommand{\eV}{\trm{\eV}}

\newcommand{\ev}[1]{\ensuremath{\langle #1 %
                     \rangle}} 
\newcommand{\tr}{\text{tr}}

\newcommand{\diag}{\text{diag}}

\begin{document}

\title{Sterile Neutrinos with Secret Interactions --- Cosmological Discord?}

\author{Xiaoyong Chu}
\email{xiaoyong.chu@oeaw.ac.at}
\affiliation{Institute of High Energy Physics, Austrian Academy of Sciences,
             Nikolsdorfergasse 18, 1050 Vienna, Austria.}

\author{Basudeb Dasgupta}
\email{bdasgupta@theory.tifr.res.in}
\affiliation{Tata Institute of Fundamental Research,
             Homi Bhabha Road, Mumbai, 400005, India.}

\author{Mona Dentler}
\email{jmodentle@uni-mainz.de}
\affiliation{PRISMA Cluster of Excellence and Mainz Institute for Theoretical Physics,
             Johannes Gutenberg University, 55099 Mainz, Germany.}

\author{Joachim Kopp}
\email{jkopp@uni-mainz.de}
\affiliation{PRISMA Cluster of Excellence and Mainz Institute for Theoretical Physics,
             Johannes Gutenberg University, 55099 Mainz, Germany.}
\affiliation{Theoretical Physics Department, CERN, 1211 Geneva, Switzerland}

\author{Ninetta Saviano}
\email{nsaviano@uni-mainz.de}
\affiliation{PRISMA Cluster of Excellence and Mainz Institute for Theoretical Physics,
             Johannes Gutenberg University, 55099 Mainz, Germany.}

\preprint{TIFR/TH/18-xxxx}

\begin{abstract}
Several long-standing anomalies from short-baseline neutrino oscillation
experiments -- most recently corroborated by new data from MiniBooNE -- have
led to the hypothesis that extra, ``sterile'', neutrino species might exist.
Models of this type face severe cosmological constraints, and several ideas
have been proposed to avoid these constraints.  Among the most widely discussed
ones are models with so-called ``secret interactions'' in the neutrino sector.
In these models, sterile neutrinos are hypothesized to couple to a new
interaction, which dynamically suppresses their production in the early
Universe through finite-temperature effects.  Recently, it has been argued that
the original calculations demonstrating the viability of this scenario need to
be refined.  Here, we update our earlier results from \emph{arXiv:1310.6337
[JCAP 1510 (2015) no.10, 011]} accordingly. We confirm that much of the
previously open parameter space for secret interactions is in fact ruled out by
cosmological constraints on the sum of neutrino masses and on free-streaming of
active neutrinos.  We then discuss possible modifications of the vanilla
scenario that would reconcile sterile neutrinos with cosmology.
\end{abstract}

\maketitle

\section{Introduction}
\label{sec:intro}

A quote {famously} attributed to Isaac Asimov is ``the most exciting phrase
to hear in science, the one that heralds new discoveries, is not `Eureka' but
\mbox{`That's funny \dots'\;}''.  Neutrino physics is arguably a field of research
where this phrase can be heard rather frequently.  Currently, it applies for
instance to several independent anomalies observed in short baseline neutrino
oscillation experiments~\cite{Aguilar:2001ty, AguilarArevalo:2012va,
Mueller:2011nm, Mention:2011rk, Hayes:2013wra, Acero:2007su}. Most recently,
interest in these anomalies has been renewed when new data from the MiniBooNE
experiment at Fermilab corroborated its earlier results
\cite{Aguilar-Arevalo:2018gpe}.  The anomalies have been
interpreted as possible hints for the existence of a fourth (``sterile'')
neutrino flavor, even though global fits indicate that it is not possible to
interpret \emph{all} experimental results in such a
scenario~\cite{Kopp:2011qd, Kopp:2013vaa, Conrad:2012qt, Kristiansen:2013mza,
Giunti:2013aea, Collin:2016aqd, Capozzi:2016vac, Dentler:2017tkw,
Dentler:2018sju}. This conclusion remains true in scenarios with more than
one sterile neutrino~\cite{Conrad:2012qt, Kopp:2013vaa}.
However, the possibility remains that some anomalies are heralding new
physics while others have mundane explanations.  Even more interesting
would be the possibility that the new physics is richer than just a sterile
neutrino (see for instance \cite{Bai:2015ztj, Liao:2016reh}).
In any case, a very severe
trial that sterile neutrino models must face is that of cosmology.  More
specifically, observations of the cosmic microwave background (CMB), of light
element abundances from Big Bang Nucleosynthesis (BBN), and of large scale
structures (LSS) in the Universe constrain the total energy in relativistic
species, usually expressed in terms of the effective number of neutrino
species, $N_\text{eff}$~\cite{Hamann:2011ge, Steigman:2012ve, Ade:2015xua}.  In
addition, LSS and the CMB constrain the sum of neutrino masses,
$\sum m_\nu$~\cite{Palanque-Delabrouille:2015pga,  DiValentino:2015wba,
DiValentino:2015sam, Cuesta:2015iho}, or, more precisely, the sum of the
masses of collisionless neutrino species.

However, cosmology can only probe particle species that are abundant in the
early Universe. It is therefore interesting to explore scenarios where sterile
neutrinos, in spite of having $\mathcal{O}(10\%)$ mixing with the active
neutrinos, are not produced in sufficient abundance to have observable
consequences.  One proposed mechanism to achieve this is the ``secret
interactions'' scenario~\cite{Hannestad:2013ana, Dasgupta:2013zpn}, in which
sterile neutrinos, while being singlets under the SM gauge group, are coupled
to a new $U(1)'$ gauge boson $A'$ (or to a new
pseudoscalar~\cite{Archidiacono:2014nda, Archidiacono:2015oma,
Archidiacono:2016kkh}) with mass $M \ll M_W$. Through this new interaction,
sterile neutrinos feel a new temperature-dependent matter potential, which
dynamically suppresses their mixing with active neutrinos at high temperatures,
while being negligible today. To avoid constraints on $N_\text{eff}$, it is in
particular required that active--sterile neutrino mixing is strongly suppressed
at temperatures $\gtrsim \text{MeV}$, the temperature where active neutrinos
decouple from the photon bath.  Note that introducing new interactions
in the sterile neutrino sector may also be one way of reconciling
the LSND and MiniBooNE anomalies with other neutrino oscillation data
\cite{Bai:2015ztj}.

While the secret interactions scenario has motivated a number of model building
and phenomenology papers~\cite{Archidiacono:2013dua, Bringmann:2013vra,
  Archidiacono:2014nda, Ng:2014pca, Kopp:2014fha, Saviano:2014esa,
  Mirizzi:2014ama, Cherry:2014xra, Kouvaris:2014uoa, Tang:2014yla, Chu:2015ipa,
Archidiacono:2016kkh, Cherry:2016jol, Forastieri:2017oma, Song:2018zyl}, it has
also been argued that most of the available parameter space is ruled out.
Constraints come mainly from two directions.  First, sterile neutrinos will
eventually recouple with active neutrinos and are then efficiently produced
collisionally via the Dodelson--Widrow mechanism~\cite{Dodelson:1993je,
Saviano:2014esa}.  The temperature at which this recoupling happens depends on
the interplay of the effective potential that suppresses flavor-changing
collisions and the relevant scattering rates, which can be very large (see
below) \cite{Cherry:2016jol}.  Even if recoupling happens at $T \ll
\text{MeV}$, it will still lead to equilibration between active and sterile
neutrinos. This may lead to tension with limits on $\sum
m_\nu$~\cite{Mirizzi:2014ama}.  Second, mixing of active and sterile neutrinos
leads to reduced free-streaming of active neutrinos. A certain amount of active
neutrino free-streaming is, however, required by CMB
observations~\cite{Archidiacono:2013dua, Forastieri:2017oma}.  Note that this
second constraint spoils attempts to keep sterile neutrinos safe from the
limit on $\sum m_\nu$ by postulating that they interact so strongly that they
cannot free-stream enough to affect large scale structure~\cite{Chu:2015ipa}.

In this paper, we take two more steps in the ongoing exploration of secretly
interacting sterile neutrinos. First, we update our earlier results from
ref.~\cite{Chu:2015ipa}, confirming in particular the findings by Cherry et
al.~\cite{Cherry:2016jol}. A detailed account of cosmological constraints on
secret interactions has also been given recently in \cite{Song:2018zyl}. In
comparison to that paper, we focus less on a complete fit to cosmological data
based on simulations, but instead derive constraints from physical
arguments and estimates that can be much more easily generalized to other models.
Second, and perhaps more importantly, we show that although the vanilla secret
interactions model is indeed disfavored by cosmological data, the general idea
underlying it remains viable and interesting.  We give explicit examples of
models that show this.  The core assumption of the secret interactions scenario
is that the sterile neutrino is hidden in cosmology because it gets a large
temperature-dependent mass at high $T$ due to its interactions.  We show that,
if the vector boson employed in the original works \cite{Hannestad:2013ana,
Dasgupta:2013zpn} is replaced by a scalar mediator with suitable symmetries and
potential, the above-mentioned constraints from BBN, CMB and LSS appear to be
avoidable.  The mechanism we propose generates a large mass for $\nu_s$ in the
high-temperature phase of the scalar potential, precluding efficient $\nu_s$
production. Only after a late phase transition in the scalar sector is the
sterile neutrino mass reduced to the value observed today.  We also outline
more mundane ways to reconcile the vanilla scenario with data, by simply adding
more free-streaming particles, or by allowing neutrinos to decay.

We begin with a review of the basic features of the secret interactions scenarios
(\cref{sec:model}), followed by the derivation of detailed cosmological
constraints (\cref{sec:constraints}).  We then discuss several possible
modifications to the original secret interactions models that could render the
scenario phenomenologically viable again (\cref{sec:reconciliation}).  We
summarize and conclude in \cref{sec:conclusions}.

\section{The Secret Interactions Scenario}
\label{sec:model}

We augment the Standard Model (SM) with an extra, sterile, neutrino flavor
$\nu_s$.  We assume $\nu_s$ has appreciable, $\mathcal{O}(10\%)$, mixing
with the three active neutrino flavors, collectively denoted by $\nu_a$.
For the neutrino mass eigenstates, we use the notation $\nu_j$, with $j=1\dots4$,
where $\nu_1$, $\nu_2$, $\nu_3$ have masses $\ll 1$\,eV and are mostly
composed of $\nu_a$.  For the mostly-sterile mass eigenstate $\nu_4$, we assume
a mass around 1\,eV, as motivated by the short baseline oscillation anomalies
\cite{Collin:2016aqd, Capozzi:2016vac, Dentler:2018sju}.  We finally introduce
the secret interaction by charging the sterile flavor eigenstate $\nu_s$
under a new $U(1)_s$ gauge group, with a gauge boson $A'$  at the MeV scale
or somewhat below.  The relevant interaction term reads
\begin{align}
  \mathcal{L}_\text{int} = e_s \bar\nu_s \gamma^\mu P_L \nu_s A'_\mu \,,
  \label{eq:Lint}
\end{align}
where $e_s$ is the $U(1)_s$ coupling constant, and \mbox{$P_L = \tfrac{1}{2} (1
- \gamma^5)$} is the projection operator onto left-chiral fermion states.  In
the following, we will be agnostic about the mechanism that breaks $U(1)_s$ and
endows the $A'$ boson with a mass.  In particular, we will neglect the possible
additional degrees of freedom---for instance sterile sector Higgs bosons---that
may be introduced to achieve this breaking.  If $A'$ gets its mass $M$ via the
St\"uckelberg mechanism, this approximation becomes exact.  When a sterile
neutrino with energy $E$ propagates through a thermalized background of sterile
neutrinos and $A'$ bosons at temperature $T_s$, it experiences a
potential~\cite{Dasgupta:2013zpn}
\begin{align}
  V_\text{eff} \simeq
  \left\lbrace 
    \begin{array}{lcl}
      -\dfrac{7 \pi^2 e_s^2 E T_s^4}{45 M^4}  &\quad& \text{for $T_s \ll M$} \\[3ex]
      +\dfrac{e_s^2 T_s^2}{8 E}               &\quad& \text{for $T_s \gg M$}
    \end{array}
  \right.\,.
  \label{eq:Veff}
\end{align}
Like a conventional Mikheyev--Smirnov--Wolfenstein (MSW)
potential~\cite{Wolfenstein:1977ue, Mikheyev:1986gs, Akhmedov:1999uz},
$V_\text{eff}$ changes the neutrino mixing angle.  In the $1+1$ flavor
approximation, the $\nu_s$--$\nu_a$ mixing angle $\theta_m$ in a thermal
$\nu_s$ background is given by
\begin{align}
  \sin^2 2\theta_m
    = \frac{\sin^2 2\theta_0}
           {\big(\cos 2\theta_0 + \tfrac{2 E}{\Delta m^2} V_\text{eff} \big)^2
           + \sin^2 2\theta_0} \,,
  \label{eq:s22thm}
\end{align}
where $\theta_0$ is the mixing angle in vacuum, and $\Delta m^2 \equiv m_4^2 - m_1^2$
is the mass squared difference between the mostly sterile and mostly active
neutrino mass eigenstates. Our qualitative results will remain unchanged even
when more than one active neutrino flavor is considered.
\Cref{eq:s22thm,eq:Veff} show that, at the high temperatures
prevalent in the early Universe, mixing is strongly suppressed because $|V_\text{eff}|
\gg \Delta m^2 / (2E)$.  Experiments today, on the other hand, will observe
$\theta_m = \theta_0$ to a very good approximation.

Note that $V_\text{eff}$ has opposite sign at $T_s \ll M$ compared to $T_s \gg
M$.  This implies that there should be a temperature range around $T_s \sim M$ where
the potential passes through zero and has a small magnitude.  In other words,
sterile neutrinos could be produced in this interval. However, as shown
explicitly in ref.~\cite{Dasgupta:2013zpn}, this temperature interval is very
short, therefore it is unclear what its impact on the final $\nu_s$ abundance
is.  Answering this question is one of the goals of this paper.

A typical cosmological history in the secret interactions model begins at high
temperature with a negligible abundance of $\nu_s$ and $A'$.  As soon as a small number
of $\nu_s$ and $A'$ are produced through oscillations or through some high-scale
interactions, a large effective potential $V_\text{eff}$ arises, suppressing
mixing and preventing further $\nu_s$ production.  When the temperature drops
so low that $|V_\text{eff}| \lesssim \Delta m^2 / (2E)$, sterile neutrinos
recouple and thermalize with $\nu_a$ through unsuppressed oscillations and
$A'$-mediated scattering processes. At $|V_\text{eff}| \simeq \Delta m^2 /
(2E)$, oscillations can even be resonantly enhanced if the recoupling
temperature is $\lesssim M$ so that $V_\text{eff}$ is negative.  If this
recoupling between $\nu_a$ and $\nu_s$ happens after the $\nu_a$ have decoupled
from the thermal bath at temperatures $\sim \text{MeV}$, $\nu_s$ production
does not change $N_\text{eff}$. The predicted value of $N_\text{eff}$ is then
similar to that in the SM, $N_\text{eff} = 3.046$~\cite{Mangano:2005cc,
deSalas:2016ztq}, and the model is consistent with the observed value
$N_\text{eff} = 3.15 \pm 0.23$ ($68\%$ CL)~\cite{Ade:2015xua}.  (The value of
$N_\text{eff}$ measured at recombination can still be reduced compared to the
SM prediction because $\nu_s$ become non-relativistic earlier than $\nu_a$.)
Nevertheless, the conversion of $\nu_a$ into $\nu_s$ increases the prediction
for $\sum m_\nu$, and for eV-scale $\nu_s$, this puts the model in tension with
the constraint $\sum m_\nu < 0.23$~eV \cite{Ade:2015xua}. Note that, to be
precise, these bounds should be slightly modified in the secret interactions
scenario as $\nu_a$--$\nu_s$ recoupling at sub-MeV temperatures lowers the
temperature of the neutrino sector compared to the standard $\Lambda$CDM
model~\cite{Mirizzi:2014ama}.  Computing this correction would require modified
simulations of structure formation, which is beyond the scope of this work.

In ref.~\cite{Chu:2015ipa}, we had proposed two possible ways out: 

\emph{(i)} Recoupling between $\nu_a$ and $\nu_s$ never happens because the
gauge coupling $e_s$ is so small that the sterile neutrino scattering rate
$\Gamma_s$ drops below the Hubble rate before $|V_\text{eff}|$ drops below
$\Delta m^2 / (2E)$. Of course, $e_s$ still needs to be large enough to
make sure that $|V_\text{eff}| \gg \Delta
m^2/(2 E)$ until active neutrino scattering decouples. However, using a
refined calculation, that we will confirm below in \cref{sec:constraints}, the
authors of ref.~\cite{Cherry:2016jol} have argued that these two contrary
requirements cannot be fulfilled simultaneously.

\emph{(ii)} Recoupling between $\nu_a$ and $\nu_s$ happens, but the gauge
coupling $e_s$ is so large that $\nu_s$ cannot free-stream until very late
times, after matter--radiation equality.  In this case, bounds on $\sum m_\nu$,
which are effectively bounds on free-streaming species, do not apply.  A
possible problem with this option is that 
the free-streaming of $\nu_a$ will be delayed as well through the $\nu_a$--$\nu_s$ mixing.
The rough estimates given in
\cite{Chu:2015ipa}, suggested that the scenario might be marginally consistent
with the data and only a dedicated analysis of CMB data would allow us to draw
definitive conclusions. Forastieri et al.\ have recently carried out such an analysis  
 and have shown that the scenario appears to be in tension with
data~\cite{Forastieri:2017oma}.

\section{Constraints on Sterile Neutrinos with Secret Interactions}
\label{sec:constraints}

In the following, we will derive updated constraints on the secret interactions
model introduced in \cref{sec:model}. We employ two complementary approaches:
our first approach, outlined in
\cref{sec:Trec}, is a computation of the recoupling temperature,
$T_\text{rec}$, i.e., the temperature at which the sterile and active neutrinos
recouple in the $1+1$ scenario.  This is similar to our previous calculations
in ref.~\cite{Chu:2015ipa}, but includes several improvements including those
pointed out in ref.~\cite{Cherry:2016jol}.  Assuming that sterile and active
neutrinos equilibrate instantaneously at $T_\text{rec}$, this computation
allows us to estimate which cosmological data sets are sensitive to the
resulting abundance of sterile neutrinos.  In the second approach, presented in
\cref{sec:sim}, we go one step further and explicitly simulate the flavor
evolution of the neutrino sector after recoupling. In doing so, we also go
beyond the $1+1$ flavor approximation and use instead a $2+1$ flavor
approximation, i.e., two active flavors and one sterile flavor. There are
several motivations to do this, as we will explain later.

\subsection{Recoupling Temperature Computation}
\label{sec:Trec}

In our first approach, we work in the mass basis and compute the production
rate $\Gamma_s$ of the mostly sterile mass eigenstate. The following reactions
contribute to $\nu_s \approx \nu_4$ production:
\begin{enumerate}
  \item $W$ and $Z$-mediated processes
    \begin{itemize}
      \item[\emph{(i)}] $e^- + e^+ \to \bar\nu_1 + \nu_4$ via $s$-channel
        $Z$ exchange or $t$-channel $W$ exchange;
      \item[\emph{(ii)}] $e^- + \nu_1 \to e^- + \nu_4$ via $t$-channel $Z$ exchange
        or $s$-channel $W$ exchange;
      \item[\emph{(iii)}] $e^+ + \nu_1 \to e^+ + \nu_4$ via $t$-channel $Z$ exchange
        or $s$-channel $W$ exchange;
      \item[\emph{(iv)}] $\nu_1 + \nu_1 \to \nu_1 + \nu_4$ via $Z$ exchange in the $t$- or
        $u$-channel;
      \item[\emph{(v)}] $\bar\nu_1 + \nu_1 \to \bar\nu_1 + \nu_4$ via $Z$ exchange in
        the $s$- or $t$-channel;
    \end{itemize}

  \item $A'$-mediated processes
    \begin{itemize}
      \item[\emph{(vi)}] $\bar\nu_4 + \nu_1 \to \bar\nu_4 + \nu_4$ via $A'$ exchange in
        the $s$- or $t$-channel;
      \item[\emph{(vii)}] $\nu_4 + \nu_1 \to \nu_4 + \nu_4$ via $A'$ exchange in the
        $t$- or $u$-channel;
    \end{itemize}
\end{enumerate}
Of course, the corresponding $CP$-conjugate processes contribute equally.
Analytical expressions for the cross sections of these reactions are given in the appendix.
Of the processes listed here, the first five are SM reactions
involving electrons and/or electron neutrinos that produce sterile neutrinos
through the mixing of the light neutrinos with the heavy mass eigenstate
$\nu_4$. The remaining two processes are mediated by $A'$ and produce sterile
neutrinos from electron neutrinos through the overlap of $\nu_e$ and $\nu_4$.
Note that process \emph{(vi)}, $\bar\nu_4 + \nu_1 \to \bar\nu_4 + \nu_4$,
involves $s$-channel $A'$ exchange and is thus resonantly enhanced in a
specific part of the neutrino spectrum. Note also that $A'$-mediated
$t$-channel scattering is enhanced in the forward direction if the $A'$ mass is
much smaller than the neutrino temperature.

We first compute the temperature $T_\text{rec}$ at which sterile and active
neutrinos recouple via scattering.  We define $T_\text{rec}$ as the temperature
at which $\Gamma_s$ becomes equal to the Hubble rate, i.e., $\Gamma_s = H$.  In
terms of the scattering cross sections given in the appendix~\ref{sec:appendix}, $\Gamma_s$ is
given by
\begin{align}
  \Gamma_s &= c_{QZ} \Big[
                    \ev{\sigma v}_{ee \to 14} \frac{n_e^2}{n_\nu}
              \;+\; \ev{\sigma v}_{e1 \to e4} \, n_e
                                                        \nonumber\\[0.2cm]
              &\quad\quad\,
               +     \ev{\sigma v}_{11 \to 14} \, {n_\nu}
               \;+\; \ev{\sigma v}_{14 \to 44} \,{n_s}
             \Big],
  \label{eq:Gamma-s}
\end{align}
Here, the notation $\ev{\cdot}$ refers to averaging over the momentum
distributions of the involved particles.  We assume these distribution
to have a Fermi-Dirac form at all times. Note also that all cross sections
depend on the the sterile sector temperature $T_s$ through the
mixing angle $\theta_m$.
The shorthand notation $\ev{\sigma v}_{ee \to 14}$ refers to process
\emph{(i)} above, $\ev{\sigma v}_{e1 \to e4}$ refers to the sum of processes
\emph{(ii)} and \emph{(iii)}, $\ev{\sigma v}_{11 \to 14}$ to the sum of
processes \emph{(iv)} (multiplied by a factor $1/2$ to account for the
identical particles in the initial state)
and \emph{(v)}, and $\ev{\sigma v}_{14 \to 44}$ to the
sum of processes \emph{(vi)} and \emph{(vii)}. The factors $n_e$, $n_\nu$,
and $n_s$ are the electron, active neutrino, and sterile neutrino number densities,
respectively, not including their anti-particles.
They are chosen such that each term in \cref{eq:Gamma-s} gives
the production rate per active neutrino, i.e.,\ the number of sterile
neutrinos produced per unit time in a spatial volume element occupied on average
by one active neutrino.
The prefactor $c_{QZ}$ accounts for the quantum Zeno effect, i.e.,
for the suppression of $\nu_s$ production when the scattering rate is faster
than the oscillation frequency~\cite{Harris:1980zi}.  In this case, oscillations have no time to
develop before they are interrupted by scattering. To account for this
effect, we define $c_{QZ}$ as
\begin{align}
  c_{QZ} = \frac{(L^\text{scat} / L^\text{osc})^2}
                {1 + (L^\text{scat} / L^\text{osc})^2} \,,
  \label{eq:QZ}
\end{align}
where $L^\text{scat}$ is the $\nu_s$--$\nu_s$ scattering length and
$L^\text{osc}$ is the oscillation length in medium.  With this definition,
$c_{QZ}$ is close to one when $L^\text{scat} \gg L^\text{osc}$ and
approaches zero when $L^\text{scat} \ll L^\text{osc}$.

\subsection{Multi-flavor evolution}
\label{sec:sim}

To understand the dynamics of sterile neutrino production in more detail, we
have also simulated the evolution of a $2+1$ system (two active species
and one sterile species) numerically.  We do so, \emph{(i)} to verify that thermalization
between active and sterile neutrinos is indeed quasi-instantaneous after
recoupling, \emph{(ii)} to assess the impact of a nearly vanishing 
$V_\text{eff}$ at $T_s \sim M$, \emph{(iii)} to check that the simplified
treatment of the quantum Zeno correction in \cref{sec:Trec} is valid, and
\emph{(iv)} to investigate the possible impact of going beyond the two-flavor
approximation.

As a complete numerical simulation of the flavor evolution including the
exact temperature-dependence of $V_\text{eff}$ is numerically highly challenging,
we focus on the evolution during the epochs where the effective potential is small
compared to the vacuum oscillation frequency, so that sterile neutrino
mixing is unsuppressed.  We use the exact temperature-dependence of
$V_\text{eff}$ from ref.~\cite{Dasgupta:2013zpn} to determine the relevant
temperature intervals, and then simulate the flavor evolution within
these intervals, setting $V_\text{eff} = 0$.  Our simulation code is
based on refs.~\cite{Saviano:2014esa, Mirizzi:2014ama, Forastieri:2017oma}.

\begin{figure*}
  \begin{center}
  \begin{tabular}{c@{\qquad}c}
    \includegraphics[width=0.4\textwidth]{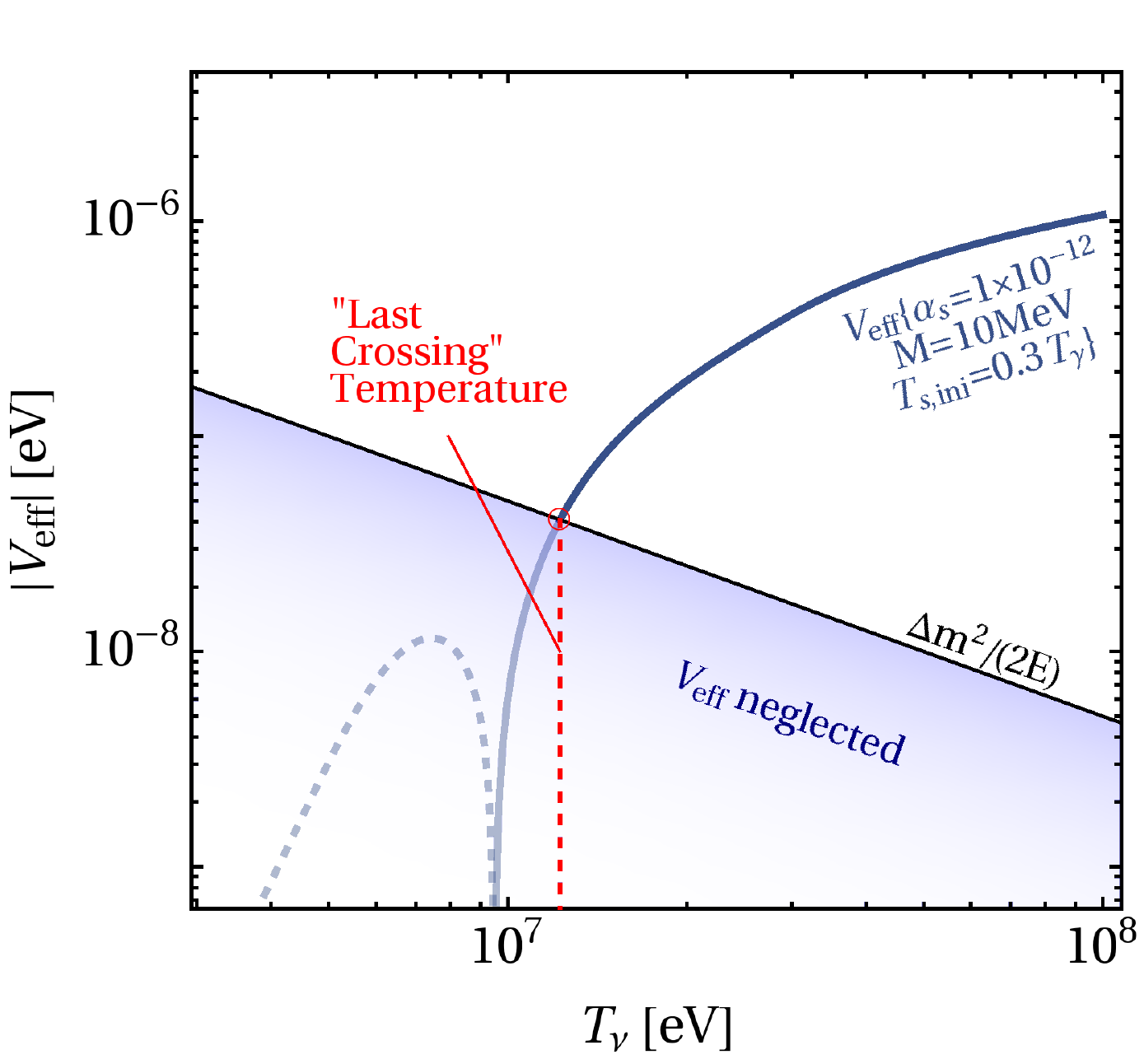} &
    \includegraphics[width=0.4\textwidth]{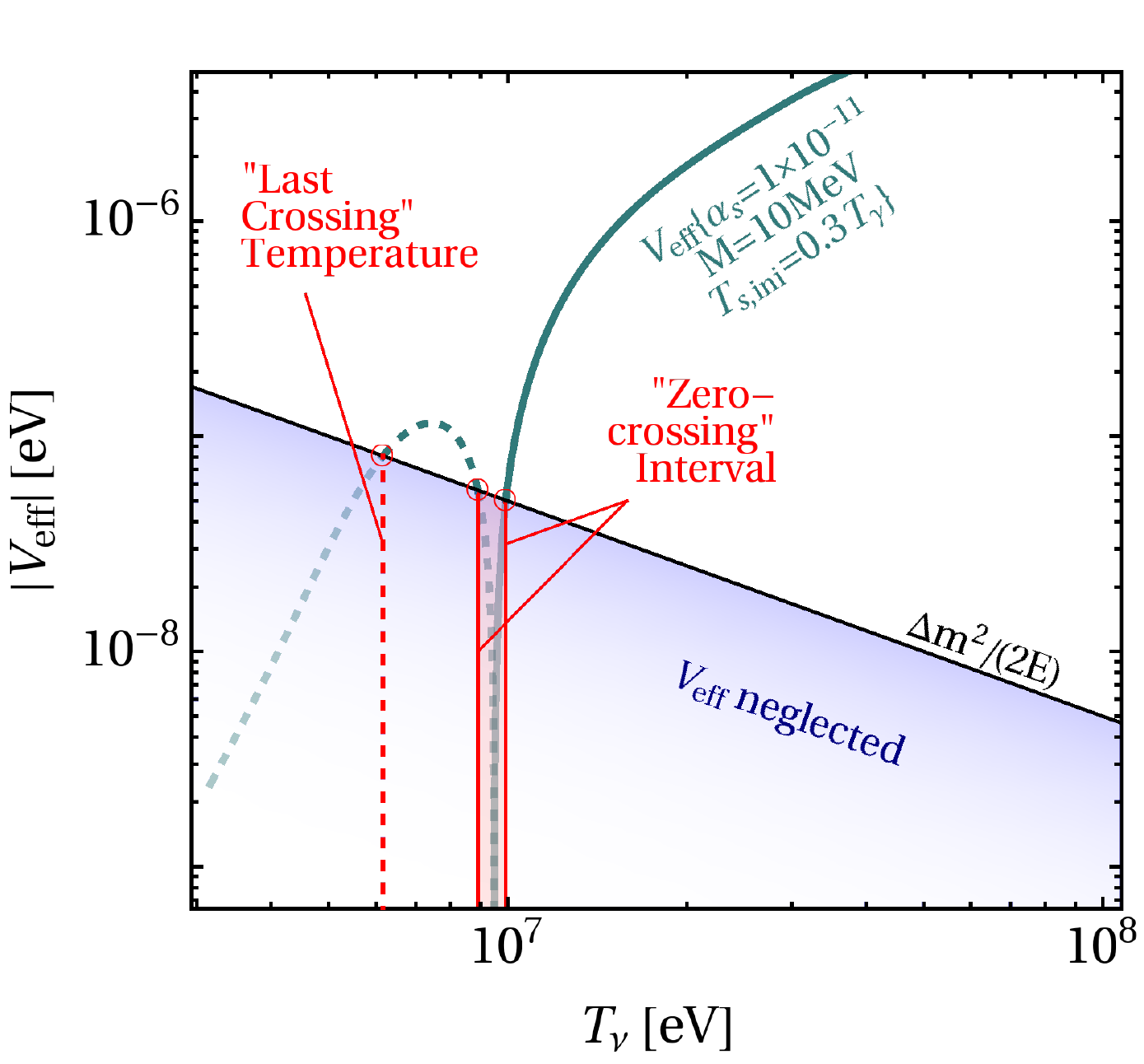}
  \end{tabular}
  \end{center}\vspace{-.4cm}
  \caption{Absolute value of the effective potential $V_\text{eff}$ as a 
    function of the (active) neutrino temperature $T_\nu$,
    for two different choices of the mediator mass $M$ and the secret fine
    structure constant $\alpha_s\equiv e_s^2/(4\pi)$.
    Positive (negative) values of $V_\text{eff}$
    are indicated by solid (dashed) lines. The vacuum oscillation frequency
    $\Delta m^2 / (2E)$ is displayed as a black line. The temperatures at which
    $|V_\text{eff}|$ and the vacuum oscillation frequency intersect are highlighted
    in red. We refer to the temperature of the last (left-most) intersection as
    the ``last crossing'' temperature. For some choices of 
    $M$ and $\alpha_s$, intersections between $|V_\text{eff}|$ and $\Delta m^2 /
    (2E)$ also occur around the temperature where  $V_\text{eff}$ changes sign,
    as can be observed in the right panel. In this case, we refer to the short
    time interval in which $|V_\text{eff}| < \Delta m^2 / (2E)$
    as the ``zero-crossing'' interval. In our simulations, we assume no sterile neutrino
    production when $|V_\text{eff}| > \Delta m^2 / (2E)$, and we set
    $V_\text{eff} = 0$ whenever $|V_\text{eff}| > \Delta m^2 / (2E)$.}
  \label{fig:Veff}
\end{figure*}

The effective potential for a sterile neutrino with 4-momentum $k$ is given by
\cite{Dasgupta:2013zpn}
\begin{align}  
  V_\text{eff} = 
    -\frac{1}{2\vec{k}^2} \Big[ \big[(k^0)^2 - \vec{k}^2]
     \tr\big( \slashed{u} \Sigma(k) \big)
    - k^0 \tr\big( \slashed{k} \Sigma(k) \big) \Big] \,,
  \label{eq:Veff_exact}
\end{align}
with $\Sigma(k)$ the temperature-dependent sterile neutrino self energy at one-loop
and $u = (1,0,0,0)$ the 4-momentum of the heat bath.
We use the ultra-relativistic approximation $k^0 \simeq| \vec{k}| +
V_\text{eff}$ and expand \cref{eq:Veff_exact} in $V_\text{eff}$. We can then
solve numerically for the critical points where the condition $|V_\text{eff}| =
\Delta m^2 / (2E)$ is fulfilled. 

At high enough temperatures, $|V_\text{eff}|$ always exceeds $\Delta m^2 /
(2E)$ as long as the fine structure constant $\alpha_s$ $(\equiv e_s^2/4\pi)$
is not zero, but as temperatures become smaller two
possibilities present themselves.  The first possibility is that once
$|V_\text{eff}|$ falls below $\Delta m^2 / (2E)$, it never exceeds it again. An
example of this is shown in the left panel of \cref{fig:Veff}. The second
possibility is that $V_\text{eff}$ crosses through zero but then takes large
negative values so that $|V_\text{eff}|$ exceeds $\Delta m^2 / (2E)$ again, as shown in the
right panel of \cref{fig:Veff}.  We refer to the temperature at which
$|V_\text{eff}|$ intersects the vacuum term for the last time as the ``last
crossing'' temperature.  In the second scenario, $|V_\text{eff}|$ intersects
the vacuum term around the zero crossing as well (\cref{fig:Veff} right), and
we call the corresponding temperature interval the ``zero-crossing'' interval.

We describe the neutrino ensembles  in terms of a
momentum-integrated $3 \times 3$ matrix of densities, 
\begin{align}
  \rho = \begin{pmatrix}
                      \rho_{ee} &  \rho_{e \mu} & \rho_{es} \\
                      \rho_{\mu e}  & \rho_{\mu \mu} &  \rho_{\mu s} \\
                      \rho_{s e} &\rho_{s \mu} &\rho_{ss}
                    \end{pmatrix} \,, 
  \label{eq:rho}
\end{align}
and a similar expression for antineutrinos, denoted by $\bar\rho$.  The
diagonal entries are the respective number densities, while the off-diagonal
ones encode phase information and vanish for zero mixing.  In the standard
situation, the equilibrium initial condition for the active neutrino number
densities is  $\rho_{ee} = \rho_{\mu\mu} = 1$ (and similarly for
$\bar\rho$), while for the sterile species we have the initial condition
$\rho_{ss} = \bar\rho_{ss} \simeq  0$.  The normalization of $\rho$ and
$\bar\rho$ is chosen such that a diagonal entry of 1 corresponds to the
abundance of a single neutrino (or antineutrino) species in the Standard Model.

The evolution equation for $\rho$ is~\cite{Sigl:1992fn,McKellar:1992ja,Mirizzi:2012we}
\begin{align}
  i \frac{d\rho}{dt} = [{\sf\Omega},\rho] + C[\rho] \,.
  \label{eq:drhodt}
\end{align}
Once again, a similar equation holds for the antineutrino density matrix
$\bar\rho$.  Here, $t$ is the comoving observer's proper time.  The evolution
can be easily recast into a function of the photon temperature $T_\gamma$.
The first term on the right-hand side of \cref{eq:drhodt} describes flavor
oscillations, with the Hamiltonian given by
\begin{align}
  {\sf\Omega} = U^\dag \bigg\langle \frac{\text{m}^2_\nu}{2 p} U \bigg\rangle
    + \sqrt{2}\,G_{\rm F} \bigg[ -\frac{8 \ev{p}}{3} \,
          \bigg( \frac{{\sf E_\ell}}{M_W^2} + \frac{{\sf E_\nu}}{M_Z^2} \bigg) \bigg] \,,
\label{eq:omega}
\end{align}
where $\text{m}_\nu = \diag(m_1, m_2, m_4)$ is the neutrino mass matrix in the
mass basis, and $U$ is the $3 \times 3$ neutrino mixing matrix.  The latter
depends on three mixing angles, $\theta_{e\mu}$, $\theta_{es}$, and $\theta_{\mu s}$, using the same parameterization as in ref.~\cite{Mirizzi:2012we}.  We
take $\theta_{e\mu}$ equal to the active neutrino mixing angle
$\theta_{13}$~\cite{Capozzi:2016rtj}, and we fix the active--sterile mixing
angles and mass-squared differences at the best-fit values obtained from a
global fit to the short-baseline anomalies~\cite{Dentler:2018sju}.  The terms
proportional to the Fermi constant $G_F$ in \cref{eq:omega} encode SM matter
effects in neutrino oscillations.  In particular, the term containing
${\sf E_\ell}$ describes charged current interactions
of neutrinos with the isotropic background medium, related to the energy
density ($\propto  T_\gamma^4$) of $e^{\pm}$ pairs. The term containing ${\sf E_\nu}$
describes instead interactions of neutrinos with themselves
(self-interaction term), related to the energy
density ($\propto (\varrho + \bar\varrho) T_\nu^4$) of $\nu$ and $\bar\nu$.
Note that in both terms, it is necessary to go beyond the low energy effective
field theory of SM weak interactions (Fermi theory) and take into account
momentum-dependent corrections.  These correction terms can compete with the
leading term from pure Fermi theory because the latter is proportional to
the tiny lepton asymmetry of the Universe.
Further details are given in ref.~\cite{Mirizzi:2012we}.
We remind the reader that we set $V_\text{eff} = 0$ below the
last crossing temperature and during the zero-crossing interval.
Moreover, we also neglect the small neutrino--antineutrino asymmetry $\propto
(\varrho - \bar\varrho) T_\nu^3$.

\begin{figure*}
  \begin{center}
  \begin{tabular}{ccc}
    \includegraphics[width=0.32\textwidth]{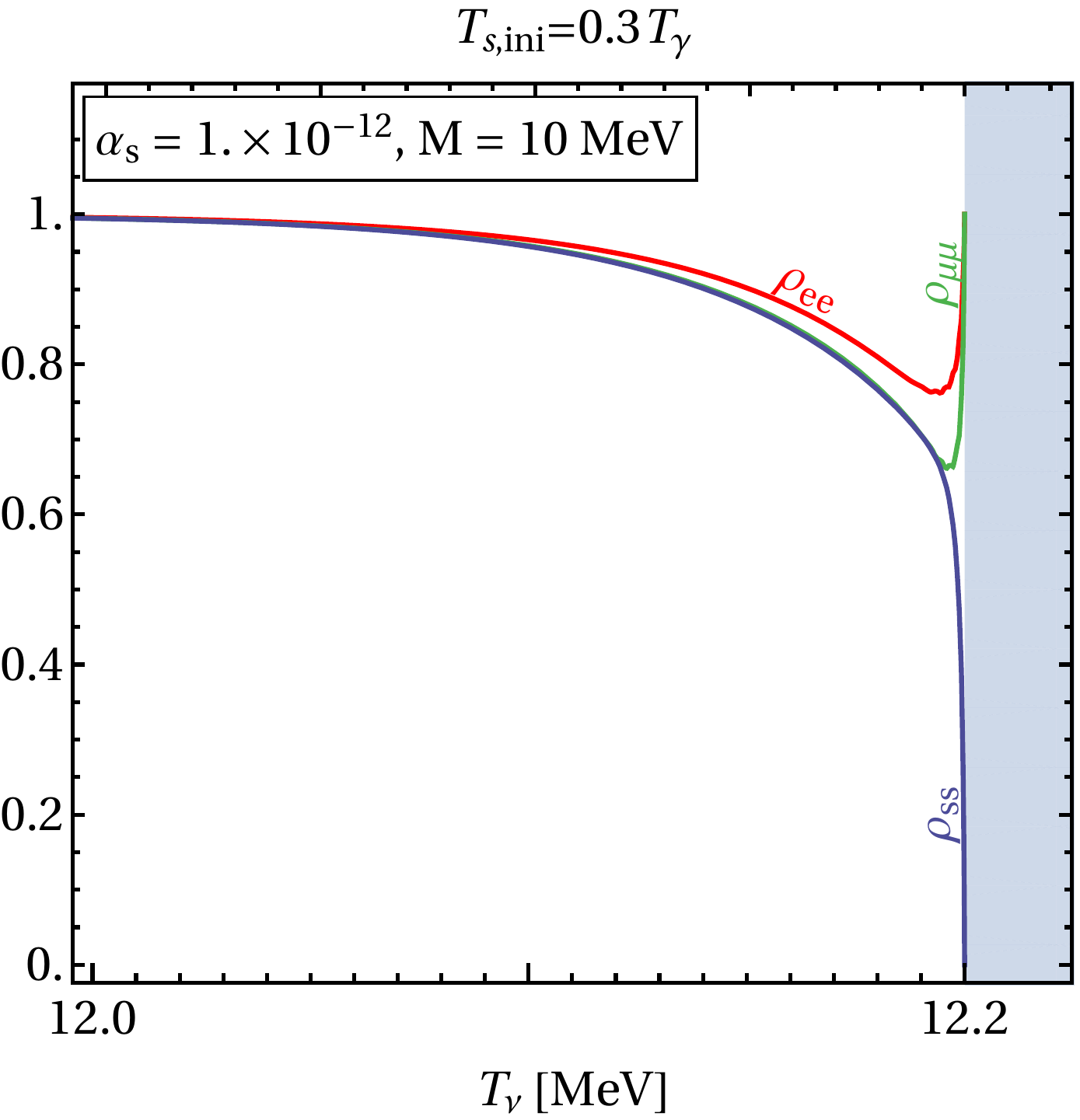} &
    \includegraphics[width=0.32\textwidth]{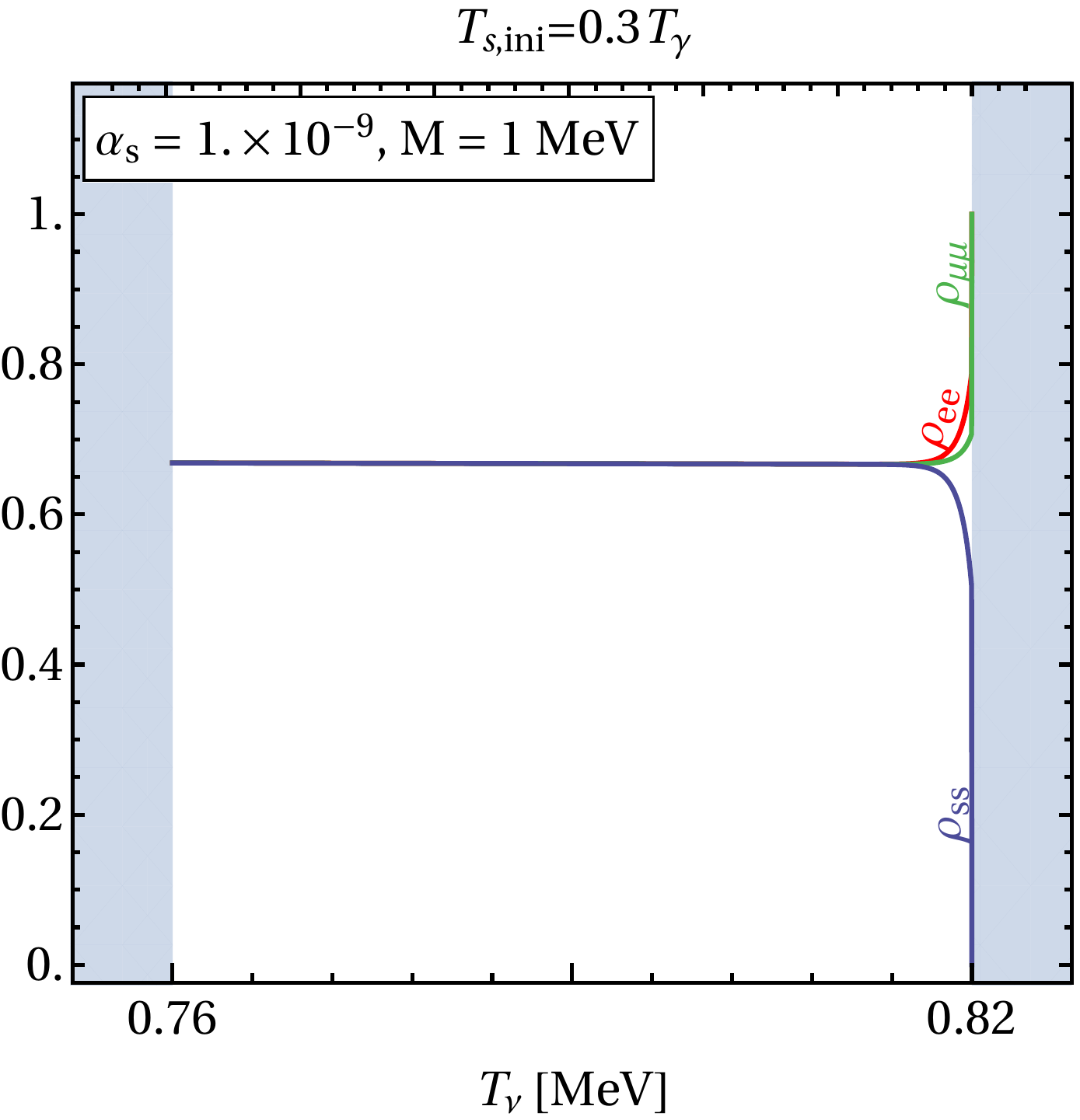} &
    \includegraphics[width=0.32\textwidth]{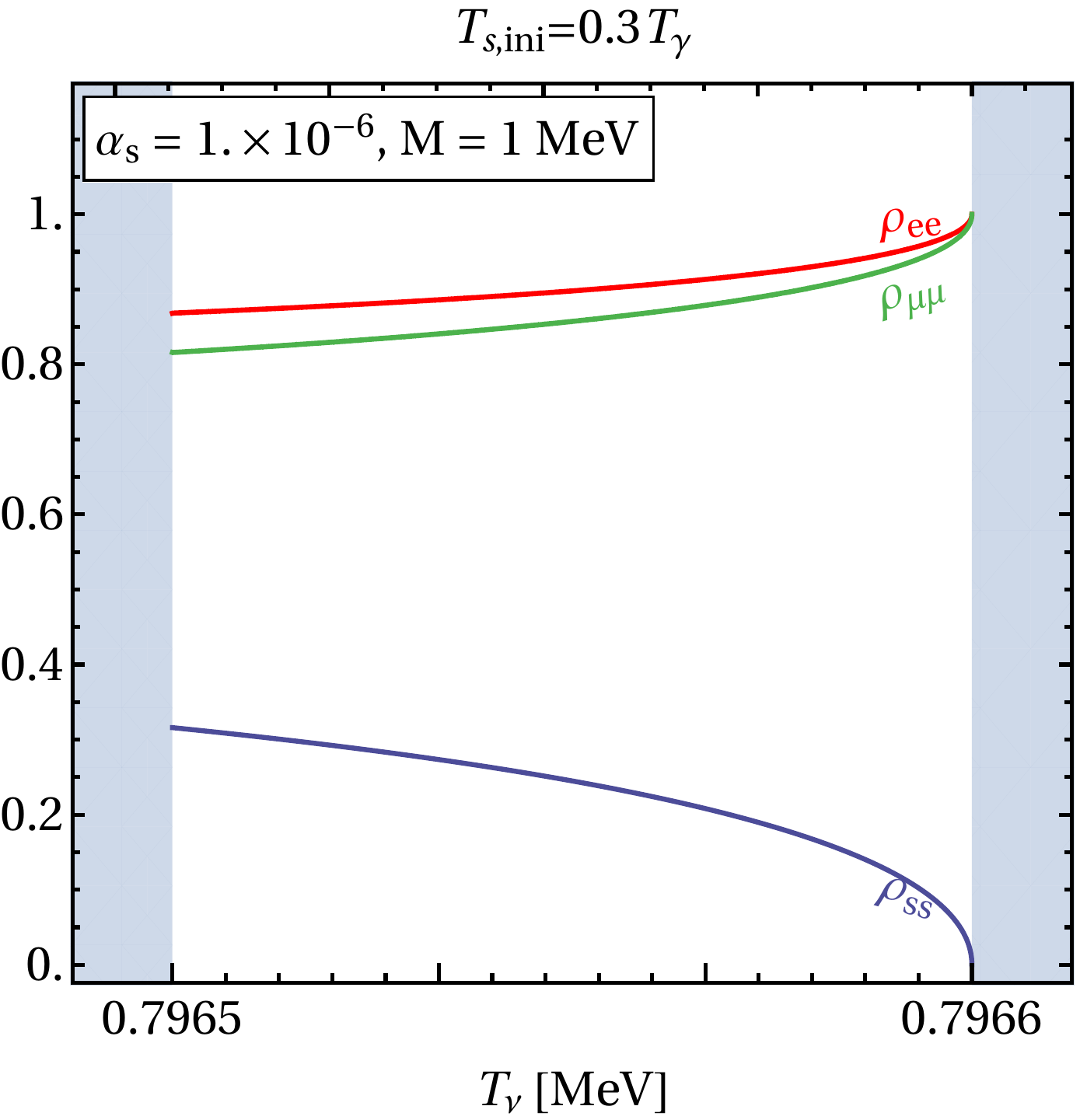} \\
    (a) & (b) & (c) 
  \end{tabular}
  \end{center}
  \label{eq:evolution}\vspace{-0.2cm}
  \caption{
    Evolution of the neutrino density matrix as a function of the (SM) neutrino
    temperature.  We show the temperature dependence of the $\nu_e$ abundance
    ($\rho_{ee}$), the $\nu_\mu$ abundance ($\rho_{\mu\mu}$), and the $\nu_s$
    abundance ($\rho_{ss}$) in the 2+1 scenario
    for three different parameter points as indicated in
    the plots.  The gray bands delimit the temperature ranges in which the system
    is evolved numerically.  Panel (a) corresponds to evolution beyond the last
    crossing temperature, panels (b) and (c) correspond to evolution within the
    zero-crossing interval.  See text for details.}
  \label{fig:multi-flavor-evol-results}
\end{figure*}

The second term on the right-hand side of \cref{eq:drhodt} is the collisional
term.  It receives contributions from both SM and secret interactions:
\begin{equation}
  C[\rho]= C_\text{SM} [\rho]+  C_{A'}[\rho] \,.
\end{equation}
Following~\cite{Mirizzi:2012we}, we write the SM collision term
as
\begin{align}
  C_\text{SM}[\rho] &=
    -\frac{i}{2} G_F^2 \, \Big[ \{{\sf S}^2, \rho - \mathbb{1}\} 
    - 2 {\sf S}(\rho - \mathbb{1}){\sf S} \nonumber \\
  & +
   \{{\sf A}^2, \rho - \mathbb{1}\} + 2 {\sf A}(\bar\rho - \mathbb{1}){\sf A} \Big] \,,
 \label{eq:collision}
\end{align}
where the matrices ${\sf S}$ and ${\sf A}$ contain the numerical coefficients
for the scattering and annihilation of the different flavors. In flavor space,
they are given by ${\sf S} = \diag(g_s^e, g_s^\mu,0)$ and 
${\sf A} = \diag(g_a^e, g_a^\mu, 0)$. Numerically one finds
\cite{Enqvist:1991qj,Chu:2006ua}: $(g_s^e)^2= 3.06$,
$(g_s^\mu)^2= 2.22$, $(g_a^e)^2= 0.5$, $(g_a^\mu)^2= 0.28$.

The collision term corresponding to secret interactions in the sterile sector
can be written schematically as
\begin{align}
  C_{A'}[\rho] &=
    -\frac{1}{2} (\Gamma_{\text{no-res}} + \Gamma_{\text{res}}) \nonumber\\
    &\qquad \times \Big[ \{{\sf S}_{A'}^2, \rho - \mathbb{1}\}
           - 2 {\sf S}_{A'}(\rho - \mathbb{1}){\sf S}_{A'} \Big] \,,
\end{align}
with the coefficient matrix ${\sf S}_{A'} \equiv \diag(0,0,1)$
\cite{Saviano:2014esa}.
Note that here, we write the scattering processes in the flavor basis,
whereas in \cref{sec:Trec} we had worked in the mass basis.  Thus,
the processes contributing to $C_{A'}$ are $\nu_s\nu_s\to\nu_s\nu_s$,                          $\nu_s\bar{\nu}_s\to\nu_s\bar{\nu}_s$, and $\nu_s\bar{\nu}_s\to A' A'$.
The coupling to $\nu_e$ is generated by the oscillation terms in the equations
of motion, but not explicitly present in $C_{A'}$.
After all, the new interaction couples only to sterile neutrinos.

To highlight the qualitative differences between different contributions to
the collision term, we have artificially split $C_{A'}$ into a piece
containing non-resonant scattering processes (including $t$-channel processes)
and a piece containing the contribution from resonantly enhanced                               $\nu_s\bar{\nu}_s\to\nu_s\bar{\nu}_s$ scattering through $s$-channel $A'$
exchange. The former piece contains the scattering rate
\begin{align}
  \Gamma_{\text{no-res}}\simeq  \frac{16\pi^2\alpha_s^2 T_\nu^5}{T_\nu^2 M^2+M^4} \,,
\end{align}
while the latter one contains
\begin{align}
  \Gamma_{\text{res}}
    \simeq  \frac{M}{T_\nu} \cdot n_s^{\text{res}} \cdot  \sigma_\text{CM} \cdot v
    \simeq  6\times 10^{-2} \alpha_s \frac{M^2}{T_\nu} \,.
\end{align}
Note that, at $T \ll M$, $\Gamma_\text{res}$ would receive an extra Boltzmann
suppression factor. In these expressions, we omit $\mathcal{O}(1)$ numerical
prefactors for simplicity.  We use the notation $n_s^\text{res} = 0.06 \, T_\nu
M \Gamma_{A'}$ for the number density of neutrinos participating in the
resonant $s$-channel process, i.e.\ particles whose energies fall within the
resonance window of width $\Gamma_{A'}=\alpha_s M/3\pi $ (in the center of mass
frame); $\sigma_\text{CM} = \pi/ M^2$ is the $s$-channel cross section  in the
center of mass frame, and $v = M/T_\nu$ is the relative velocity of the two
neutrinos forming a resonant pair.

In \cref{fig:multi-flavor-evol-results} we show the results of the numerical
flavor evolution for sterile and active neutrinos at three representative
parameter points.  In particular, we show the evolution of the density matrix
components $\rho_{ee}$ (electron neutrino abundance relative to a fully
thermalized species), $\rho_{\mu\mu}$ (muon neutrino abundance), and
$\rho_{ss}$ (sterile neutrino abundance). Panel (a), where $M = 10$~MeV, $\alpha_s =
10^{-12}$ was assumed, corresponds to the evolution after the last crossing
temperature, while panel (b), with $M =1$~MeV, $\alpha_s = 10^{-9}$, and panel
(c), with $M = 1$~MeV, $\alpha_s = 10^{-6}$, correspond to the evolution during
the zero crossing interval. In computing the zero-crossing temperatures and
the last crossing temperature, we need to make an assumption on the initial
temperature $T_{s,\text{ini}}$, before any $\nu_s$ are produced via
oscillations.  Here, we assume $T_{s,\text{ini}} = 0.3 T_\gamma$ at $T_\gamma
= 1$~TeV.  We will motivate this choice below in \cref{sec:results}.
Our assumptions on $T_{s,\text{ini}}$ and on the evolution of $T_s$ are of
course irrelevant to the actual numerical evolution of the neutrino ensemble
as we set $V_\text{eff} = 0$ in the zero-crossing interval and after the last
crossing temperature.  The grey bands in
\cref{fig:multi-flavor-evol-results} delimit the temperature range during which
we perform the numerical flavor evolution. Within the gray bands,
$V_\text{eff}$ cannot be considered negligible any more.  In the cases shown in
panels (a) and (b), sterile neutrinos are copiously produced. In particular, in
panel (a), sterile neutrinos are fully thermalized ($\rho_{ss}=1$) by
oscillations with active neutrinos occurring at temperatures around 10~MeV and
increasing the number of relativistic degrees of freedom $N_\text{eff}$. In
panel (b) instead, since oscillations and thus sterile neutrino production
happen after the neutrino sector has decoupled from the photon bath, the total
neutrino number density remains constant. Consequently, the asymptotic values
of $\rho_{ee}$, $\rho_{\mu\mu}$, and $\rho_{ss}$ are all 0.67 in the $2+1$
scenario.  At the parameter points shown in panels (a) and (b), the oscillation
rate is much larger than the $\nu_s$ scattering rate, i.e., there is no quantum
Zeno suppression, and the $\nu_s$ scattering rate is in turn much larger than
the Hubble rate, i.e., scattering-induced production is efficient. In panel (c)
of \cref{fig:multi-flavor-evol-results}, $\nu_s$ are not copiously  produced
during the zero-crossing interval.  The reason is that resonant scattering
mediated by an $s$-channel $A'$ is faster than vacuum oscillations so that
$\nu_s$ production is quantum Zeno-suppressed.  Note that this is not the
typical behavior -- for most combinations of $M$ and $\alpha_s$, we find
efficient $\nu_s$ production during the zero-crossing interval, except for a
few cases where the interval is very short and/or the mediator is very massive
($\sim 1$~GeV).  We always find efficient $\nu_s$ production below the last
crossing temperature.

\begin{figure*}
  \begin{center}
    \begin{tabular}{ccc}
    \includegraphics[width=0.32\textwidth]{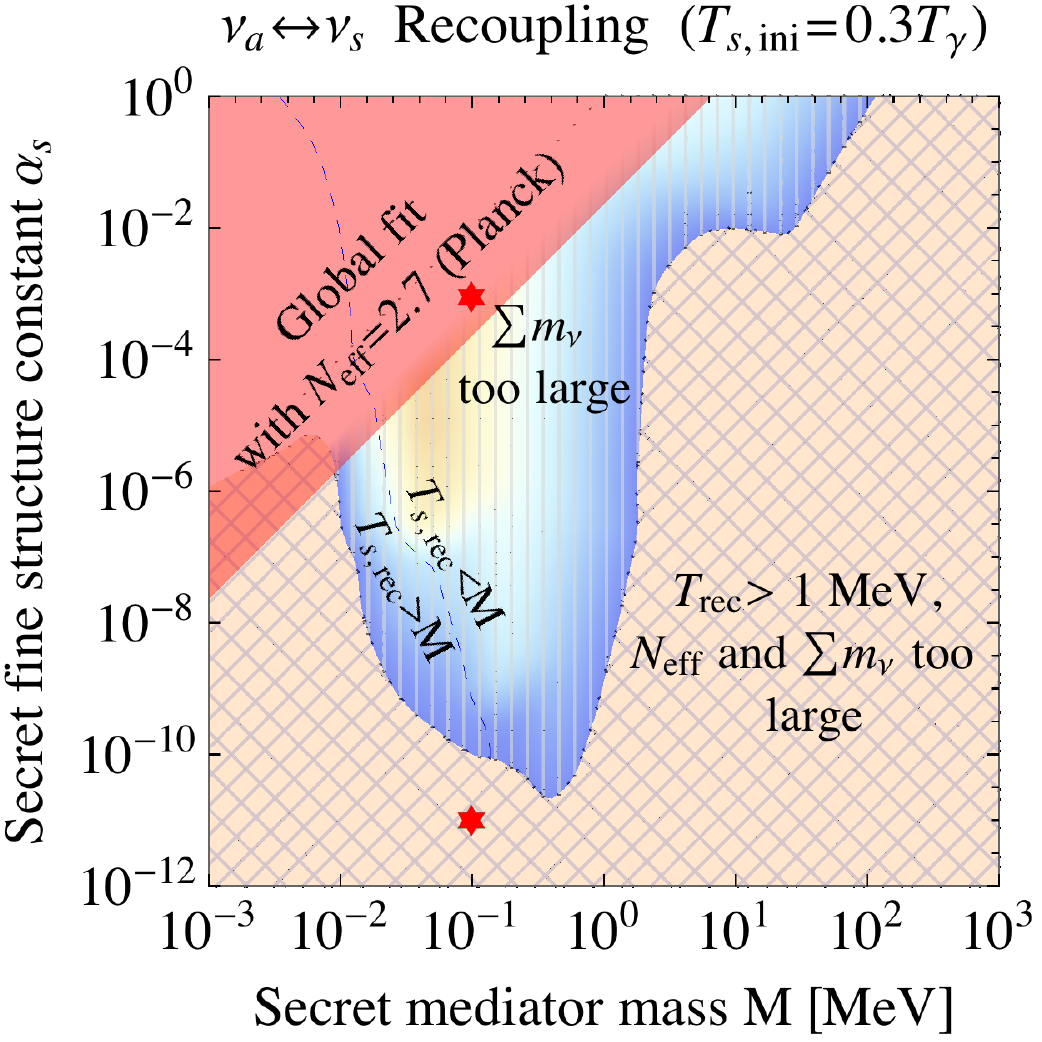} &
    \includegraphics[width=0.32\textwidth]{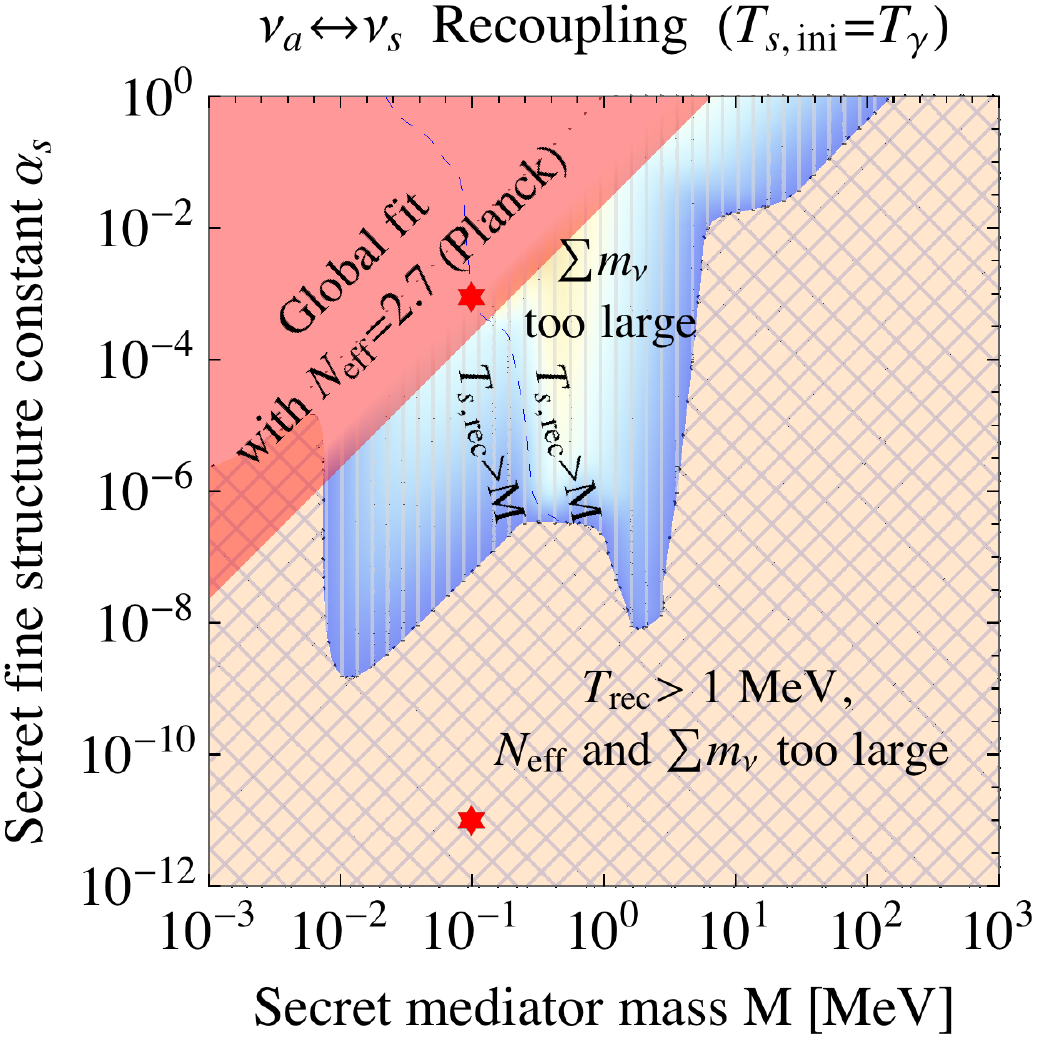} &
    \includegraphics[width=0.32\textwidth]{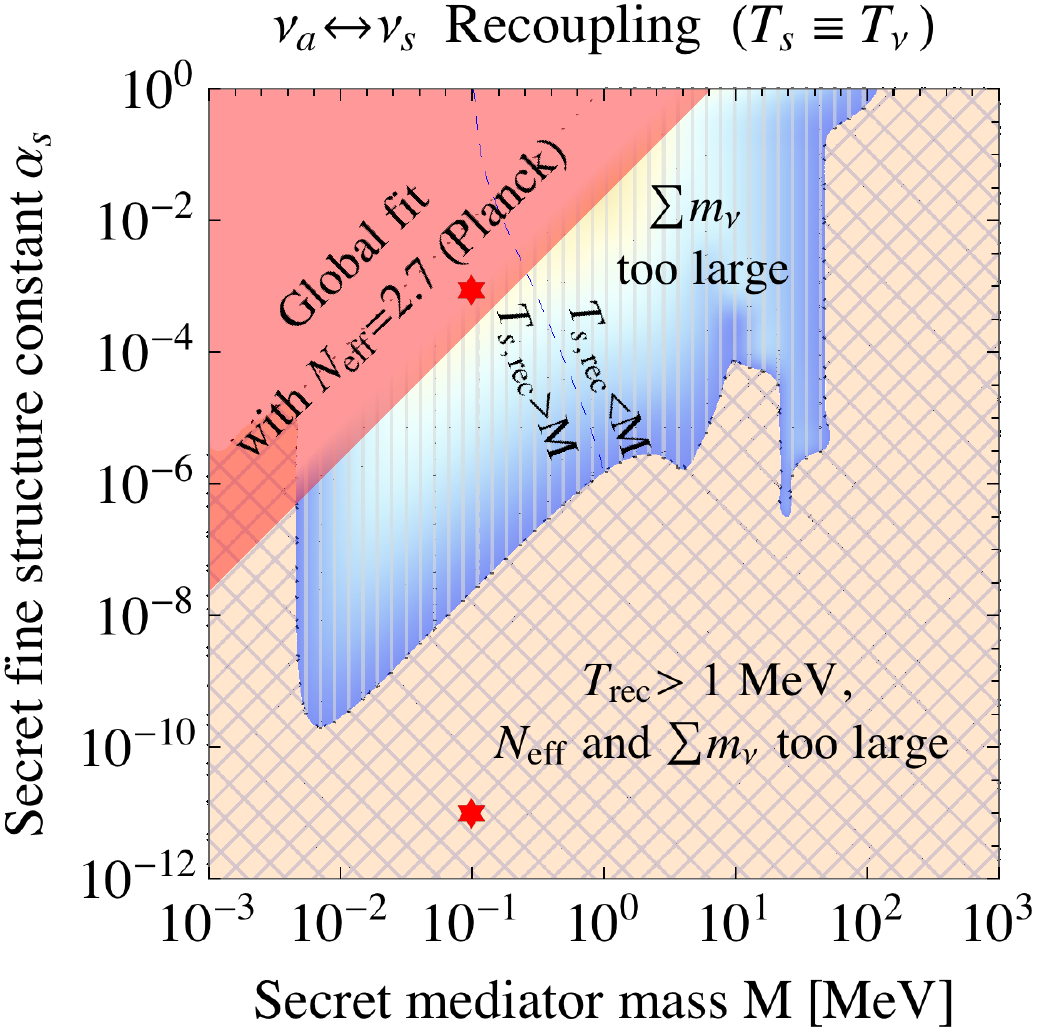}
    \end{tabular}
  \end{center}
  \vspace{-0.2cm}
  \caption{The parameter space of the secret interactions model as a function
    of the secret gauge boson mass $M$ and the corresponding fine structure
    constant $\alpha_s$, for three different assumptions on the ratio of
    the sterile and active sector temperatures (see text for details).  We have
    assumed a sterile neutrino mass $m_s = 1$\,eV, and a vacuum mixing angle
    $\theta_0 = 0.1$. The cross-hatched region (brown) is ruled out because it
    leads to recoupling at $T_\gamma > 1$~MeV, so that sterile neutrinos will
    fully thermalize with the SM plasma, in violation of the constraints on
    $N_\text{eff}$ and $\sum m_\nu$.  In the vertically striped (blue/orange)
    region, recoupling occurs at $T_\gamma < 1$~MeV, but even in this case
    $\nu_s$ are produced collisionally after $|V_\text{eff}|$ drops below
    $\Delta m^2 / (2 E)$, and the model is ruled out due to constraints on
    $\sum m_\nu$~\cite{Mirizzi:2014ama}.
    This constraint would be avoided for $m_s \lesssim 0.2$~eV.
    The color gradient in this region, from dark orange to blue, represents the
    increasing recoupling temperature from 0.05~MeV to 1~MeV.  The red shaded
    region in the top left of the figure is ruled out because of insufficient
    active neutrino free-streaming~\cite{Forastieri:2017oma}.  The red stars
    indicate two benchmark points that were considered in
    ref.~\cite{Chu:2015ipa} and are now ruled out.  There is no parameter
    region where recoupling never occurs, i.e.,\ all values of $M$ and
    $\alpha_s$ are ruled out for $m_s \geq 1$~eV.
  }
  \label{fig:paramspace}
\end{figure*}

\subsection{Results}
\label{sec:results}

In fig.\,\ref{fig:paramspace} we show the main constraints on the parameter
space of the secret interactions model in the plane spanned by the secret gauge
boson mass $M$ and the corresponding fine structure constant $\alpha_s$. We
have assumed a sterile neutrino mass $m_s = 1$\,eV, and a vacuum mixing angle
$\theta_0 = 0.1$.  We distinguish three regimes: in the vertically striped  (blue/orange) region labeled ``$\sum m_\nu$ too large'', $\nu_s$
production is efficiently suppressed down to temperatures $T_\gamma \le 1$~MeV,
so that $N_\text{eff}$ limits are evaded. Nevertheless, $\nu_s$ are efficiently
produced via collisional decoherence at late times~\cite{Mirizzi:2014ama},
i.e.,\ around or below the last crossing temperature, so that the constraint on $\sum
m_\nu$ is violated.  Note that for lighter sterile neutrinos, $m_s \lesssim
0.2$~eV, these parameter regions would be experimentally allowed. The dashed
line within the vertically hatched region indicates where the recoupling
temperature equals the $A'$ mass.  In the cross-hatched (brown) region in
\cref{fig:paramspace}, sterile neutrinos recouple above $T_\gamma \sim 1$~MeV.
They can thus fully thermalize with the SM thermal bath, and as a consequence
violate constraints on both $N_\text{eff}$ and $\sum m_\nu$.  The red shaded
region at the top left of the plots is likely ruled out by CMB data because of
insufficient active neutrino free-streaming~\cite{Forastieri:2017oma}.  The red
stars are two benchmark points considered in ref.~\cite{Chu:2015ipa}. We see
that both are now disfavored.  The boundary between the striped and
cross-hatched regions is first based on the value of $T_\text{rec}$ calculated
using the methods from \cref{sec:Trec}. These methods, however, do not properly
take into account $\nu_s$ production during the short time interval where
zero-crossing happens and after the last crossing.  Therefore, we use the
numerical simulations from \cref{sec:sim} to reexamine the zero crossing
interval and to determine whether $\nu_s$ production around the zero crossing
shifts $T_\text{rec}$ to larger values.  If so, we set $T_\text{rec}$ to the
central temperature of the zero crossing interval. We find, however, that this
correction never affects the boundary between the striped and cross-hatched regions
in \cref{fig:paramspace}.  We conclude that the sum of neutrino masses
constraint and active neutrino free-streaming constraint together rule out all
of the parameter space for the model~\cite{Forastieri:2017oma}.

The left and middle panels in \cref{fig:paramspace} correspond to different choices of
the initial temperature $T_{s,\text{ini}}$ of $\nu_s$ and $A'$ at very early
times. (We arbitrarily define $T_{s,\text{ini}}$ as the value of $T_s$ at
photon temperature $T_\gamma = 1$~TeV.) We assume that there exist some
additional new interactions between $\nu_s$ and SM particles (for instance in
the context of a Grand Unified Theory) that lead to thermalization of $\nu_s$
at a very high temperature $T_\gamma \gg \text{TeV}$. When these interactions
freeze out (still at $T_\gamma \gg \text{TeV}$), the sterile and SM sectors
decouple.  Afterwards, $T_s$ and $T_\gamma$ may  drift apart, and the amount by
which they do so above $T_\gamma = 1$~TeV is encoded in our choice of
$T_{s,\text{ini}}$.  Of course, further entropy is produced in the SM sector at
$T_\gamma < 1$~TeV, which implies that $T_s$ and $T_\gamma$ will drift further
apart as the Universe evolves. This effect is taken into account in our
calculations. Even if the sterile neutrino abundance is zero after inflation
and reheating, and sterile neutrinos are only produced via oscillations, a
non-vanishing $T_{s,\text{ini}}$ is still determined by the equation $\Gamma_s
= H$.  In other words, at any given epoch sterile neutrinos will be produced
until $V_\text{eff}$ becomes large enough to shut production off (or until full
thermal equilibrium between $\nu_s$ and $\nu_a$ is reached). In the right
panel of \cref{fig:paramspace}, we show also constraints under the hypothesis that
$T_s = T_\nu$, i.e.\ that the sterile sector temperature follows the
active neutrino temperature at all times.  This scenario, while difficult
to realize in a consistent model, can be considered an upper limit on $T_s$.

\begin{figure}
  \centering
  \includegraphics[width=0.36\textwidth]{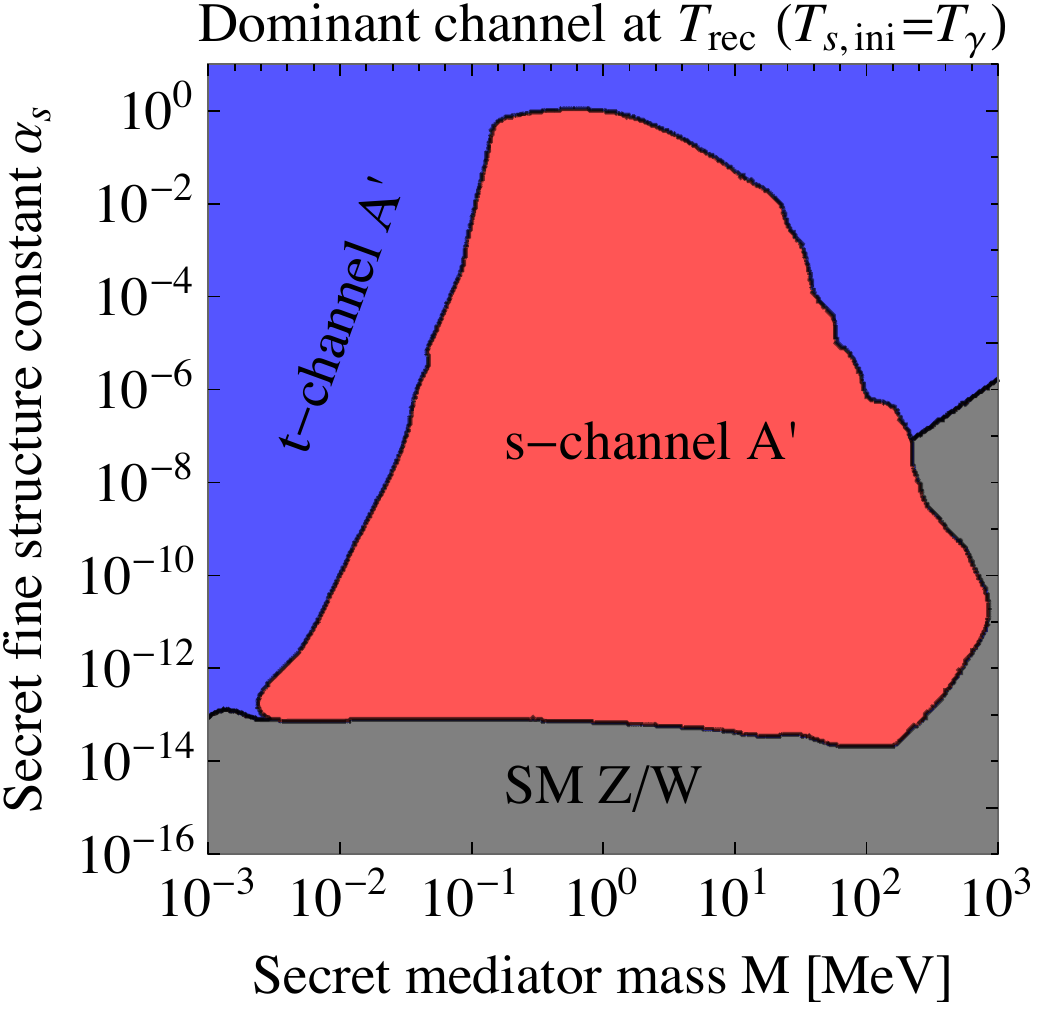}
  \caption{Dominant scattering channel for collisional $\nu_s$ production
    at $T_\text{rec}$ as a function of $M$ and $\alpha_s$.  We have chosen
    $m_s = 1$~eV, and $T_\text{s,ini} = T_\gamma$.  The mixing angle suppression
    is common to all processes.
    }
  \label{fig:dominant-channels}
\end{figure}

Among the various $\nu_s$ production processes listed above, the $W$- and
$Z$-mediated ones are dominant at the recoupling time if either $A'$ is heavy
(close to 1~GeV), or for $\alpha_s \lesssim 10^{-14}$, as shown by the gray
region in \cref{fig:dominant-channels}. The $A'$-mediated $s$-channel
contribution to process \emph{(vi)}, $\bar\nu_4 + \nu_1 \to \bar\nu_4 + \nu_4$
(shown in red), is dominant for most of the parameter region shown in
\cref{fig:dominant-channels}, largely due to the on-shell resonance.  The
$A'$-mediated $t$-channel contributions to processes \emph{(vi)} ($\bar\nu_4 +
\nu_1 \to \bar\nu_4 + \nu_4$) and \emph{(vii)} ($\nu_4 + \nu_1 \to \nu_4 +
\nu_4$), shown in blue, become more important when either the $s$-channel
resonance is Boltzmann-suppressed in the case of heavy $A'$, or when the
forward enhancement of the $t$-channel diagrams becomes significant in the case
of very light $A'$.

Note that the $A'$ resonance in the $s$-channel is responsible for ruling out
the parameter region in which it was previously thought \cite{Chu:2015ipa} that
no recoupling between $\nu_a$ and $\nu_s$ happens. The calculations  in
\cite{Chu:2015ipa} were based on naive dimensional arguments, and the
enhancements induced by on-shell resonance and forward scattering were missed.
This subtle issue was pointed out by Cherry et al. \cite{Cherry:2016jol}, who
in particular explained that while the truly forward scattering of $\nu_s$ only
gives a refractive index $V_{\rm eff}$ (which was included in previous papers),
multiple ``almost forward" small-angle scatterings, which were incorrectly
ignored,  eventually add up to give large angle scattering and spatially
separate the $\nu_1$ and $\nu_4$ eigenstates causing decoherence.

Let us finally address the potential loopholes in our line of argument so far.
We have already argued that our results -- in particular \cref{fig:paramspace} --
are unaffected by sterile neutrino production during the zero crossing
interval. We have explicitly checked this using the simulations described
in \cref{sec:sim}.  Similarly, the robustness of our results with regard to
possible corrections from the more detailed simulations also addresses the
other points raised at the beginning of \cref{sec:sim}. In particular, it
illustrates that the approximation of quasi-instantaneous thermalization
after after $V_\text{eff}$ drops below $\Delta m^2 / (2 E)$ is a good one,
that a simplified treatment of the quantum Zeno effect is usually justified,
and that there are no qualitative differences between the $1+1$ and $2+1$
scenarios.

\section{Reconciling Secret Interactions with Cosmology}
\label{sec:reconciliation}

While \cref{fig:paramspace} shows that the secret interactions model in its
vanilla form is difficult to reconcile with cosmological constraints, it is
interesting to ask what modifications are necessary to render it viable. In the
following we outline some ideas for modifying the scenario in order to
reconcile eV-scale sterile neutrinos with cosmology.  Some of the ideas
discussed here may seem rather contrived, but we discuss them nevertheless
to give the reader a feeling for what it takes to make eV-scale $\nu_s$
cosmologically viable.

We are not going to comment here on ways of reconciling sterile neutrinos with
cosmology that do not involve secret interactions.

\subsection{Recoupling never happens}
\label{sec:isb}

If sterile neutrinos are much heavier than the sterile sector temperature $T_s$
at early times and only become light when $\nu_s$-producing processes have
decoupled, their production will be suppressed.

One way to implement this idea is to use a scalar mediator $\phi$ instead of a
vector mediator $A'$.
The resulting Yukawa coupling $\phi \bar\nu_s \nu_s$ makes the mass of $\nu_s$
dependent on the vacuum expectation value (vev) $v_\phi$ of $\phi$.  If
$v_\phi$ is large at high temperatures and vanishes at lower temperatures,
$\nu_s$ can be ``hidden'' until the Universe is cold enough to suppress their
production.  Scenarios of this type have been known for a long time
\cite{Weinberg:1974hy, Bajc:1999cn, Bimonte:1999tw, Jansen:1998rj,
  Profumo:2007wc, Cohen:2008nb, Cline:2009sn, Espinosa:2011ax, Cui:2011qe,
Cline:2012hg, Fairbairn:2013uta, Curtin:2014jma, Baker:2016xzo}.

To construct a toy model based on this idea, we consider two real scalars $\phi_1$,
$\phi_2$, enjoying a $\mathbb{Z}_2 \times \mathbb{Z}_2$ symmetry under which
$\phi_1$ carries charges $(-, +)$, while $\phi_2$ carries charges $(+, -)$.
The tree level scalar potential is then
\begin{align}
  V = {\lambda_1 \over 4} \phi_1^4
    + {\lambda_2 \over 4} \phi_2^4
    + {\lambda_p \over 2} \phi_1^2 \phi_2^2
    + {\mu_1^2 \over 2} \phi_1^2
    + {\mu_2^2 \over 2} \phi_2^2 \,. 
 \label{eq:V-phi1-phi2}
\end{align}
Boundedness of the potential from below requires that $\lambda_{1,2} > 0$
and $\lambda_p^2 < \lambda_1 \lambda_2$. As long as $\mu_1^2 > 0$ and $\mu_2^2 > 0$,
there are no broken symmetries at zero temperature.

At high temperatures the potential receives thermal corrections.
At 1-loop order and with $T_s^2\gg\mu_{1,2}^2$, these are~\cite{Weinberg:1974hy}
\begin{align}
  \Delta V(T_s) &= {T_s^2 \over 24} \sum_{i=1,2} {\partial^2 V \over \partial \phi_i^2}
                                                        \nonumber\\
           &\simeq {T_s^2 \over 24} \Big[ (3\lambda_1 + \lambda_p) \phi_1^2
                                      + (3\lambda_2 + \lambda_p) \phi_2^2 \Big] \,.
\end{align}
If $3\lambda_1 + \lambda_p < 0$, the field $\phi_1$ develops a nonzero vev 
$v_{\phi 1}$ at temperatures above
\begin{align}
  T_{s,\text{crit}} \equiv T_s(T_\text{crit})
          \simeq \big[ 12 \mu_1^2 / |3\lambda_1+\lambda_p| \big]^{1/2} \,,
  \label{eq:Tsc}
\end{align}
breaking one of the $\mathbb{Z}_2$ symmetries.  Here, $T_s(T_\text{crit})$ denotes the
sterile sector temperature at the time when the photon temperature is $T_\text{crit}$.
The subscript ``crit'' stands for critical temperature.
One can see that $v_{\phi 1} \neq 0$ will occur for modestly large negative values of the
quartic cross-coupling $\lambda_p$.  Because of the boundedness conditions that
force $3\lambda_2 + \lambda_p > 0$, the other scalar $\phi_2$ cannot develop a
vev simultaneously, so one of the $\mathbb{Z}_2$ symmetries remains unbroken.
It is easy to see that if $\mu_1^2$ is very small, $T_\text{crit}$ can be quite low,
perhaps lower than the temperature $T_\text{dec}$ at which active and sterile neutrinos
finally decouple for good.

The charges of sterile neutrinos under the $\mathbb{Z}_2 \times \mathbb{Z}_2$
are chosen as follows: the left-handed component $\nu_{sL}$ of the Dirac fermion
$\nu_s$ carries charges $(+, +)$, while its right-handed partner $\nu_{sR}$
is a singlet with charges $(-,+)$. 
The couplings of $\nu_s$ are then
\begin{align}
  \mathcal{L}_\text{s} \supset
    - y \phi_1 \overline{\nu_{sL}} \nu_{sR}
    - \frac{1}{2} m_{sL} \overline{\nu_{sL}^c} \nu_{sL}
    - \frac{1}{2} m_{sR} \overline{\nu_{sR}^c} \nu_{sR} + h.c.\,.
  \label{eq:L-ISB}
\end{align}

The two main issues that we discuss now are whether this interaction is
sufficient to generate a large $v_{\phi 1}$-induced thermal mass for $\nu_s$ in
order to prohibit $\nu_s$ production until $T_\gamma < T_\text{dec}$, and
whether sterile neutrino scattering can freeze-out already at $T_\gamma > T_\text{crit}$
in order to avoid recoupling below $T_\text{crit}$.  Of course, we also have to demand
that $T_\text{crit} < 1$~MeV to prevent collisional sterile neutrino production via $Z$-
and $W$-mediated processes.

At high temperatures, $T_\gamma > T_\text{crit}$, the temperature-dependent (Dirac)
mass for $\nu_s$ is
\begin{align}
  m_s(T_s) &= y \sqrt{{-12\mu_1^2-(3\lambda_1+\lambda_p) T_s^2}\over{12\lambda_1}}
  & \text{for $T_\gamma > T_\text{crit}$} \,.
\end{align}
At $T_\gamma < T_\text{crit}$, $\nu_s$ splits into two Majorana fermions of mass $m_{sL}$
and $m_{sR}$. We assume that at least one of these is $\mathcal{O}(\text{eV})$.
To prohibit production of $\nu_s$, we demand that $m_s(T_s) \gtrsim T_s$ at
$T_s > T_{s,\text{crit}}$, so that $\nu_s$ production becomes exponentially suppressed.

As an example, if $\lambda_2 \simeq 1$ then $\lambda_1
\gtrsim \lambda_p^2$ satisfies the boundedness criterion, and $m_s(T_s) \simeq y T_s /
\sqrt{12 |\lambda_p|}$ for $-1\ll \lambda_p<0$ and $T_s \gg T_{s,\text{crit}}$.
The requirement $T_\text{crit} < T_\text{dec}$ then implies
\begin{align}
  |\lambda_p| &> 12 \mu_1^2 / T_{s,\text{dec}}^2 \,,
  \label{eq:x-condition-1}
  \intertext{while the condition $m_s(T_s) > T_s$ implies}
  |\lambda_p| &< {y^2 \over 12} \,.
  \label{eq:x-condition-2}
\end{align}

To determine $T_\text{dec}$, we need to consider $\phi_1$-mediated
neutrino--neutrino scattering. We first redefine the field $\phi_1  \to
v_{\phi 1}  + \rho $ after symmetry breaking, where the mass of the physical scalar $\rho$ is given by
$m_\rho^2(T_s) = 2 \lambda_1 v_{\phi 1}^2 \simeq T_s^2 |\lambda_p| / 6$ for
$T_\gamma > T_\text{crit}$ ($i.e.$, $T_s > T_{s,\text{crit}}$) and
$m_\rho^2 \simeq \mu_1^2 + (3 \lambda_1 + \lambda_p) T_s^2 / 12$ for
$T_\gamma \le T_\text{crit}$. 
 The decoupling temperature
is defined as the temperature at which the rate for $\nu_a$--$\nu_s$ inelastic
scattering  drops below the Hubble  rate:
\begin{align}
  n_\nu(T_\text{s,dec}) \, {\sin^2 \theta_0 \, y^4 \over m_\rho^2} = H(T_\text{dec})
                                  \simeq {T_\text{dec}^2 \over M_\text{Pl}}\,.
  \label{eq:T-dec-condition}
\end{align}
With $n_\nu(T_s) \simeq T_s^3$ and $T_\text{dec}\ge T_\text{s,dec} $, we get
\begin{align}
  T_\text{s, dec}   \ge {m_\rho^2 \over \sin^2\theta_0 \, y^4 M_\text{Pl}} \,.
  \label{eq:T-dec-solution}
\end{align}
Therefore, \cref{eq:x-condition-1,eq:x-condition-2} are true as long as
\begin{align}
  12 \sin^4\theta_0 \, y^8 \, {\mu_1^2 M_\text{Pl}^2 \over m_\rho^4}
  < |\lambda_p| < {y^2 \over 12} \,.
\end{align}
This condition can be satisfied if $y$ is tiny, and $|\lambda_p|$ is even
tinier.

\begin{figure}
  \centering
  \includegraphics[width=0.36\textwidth]{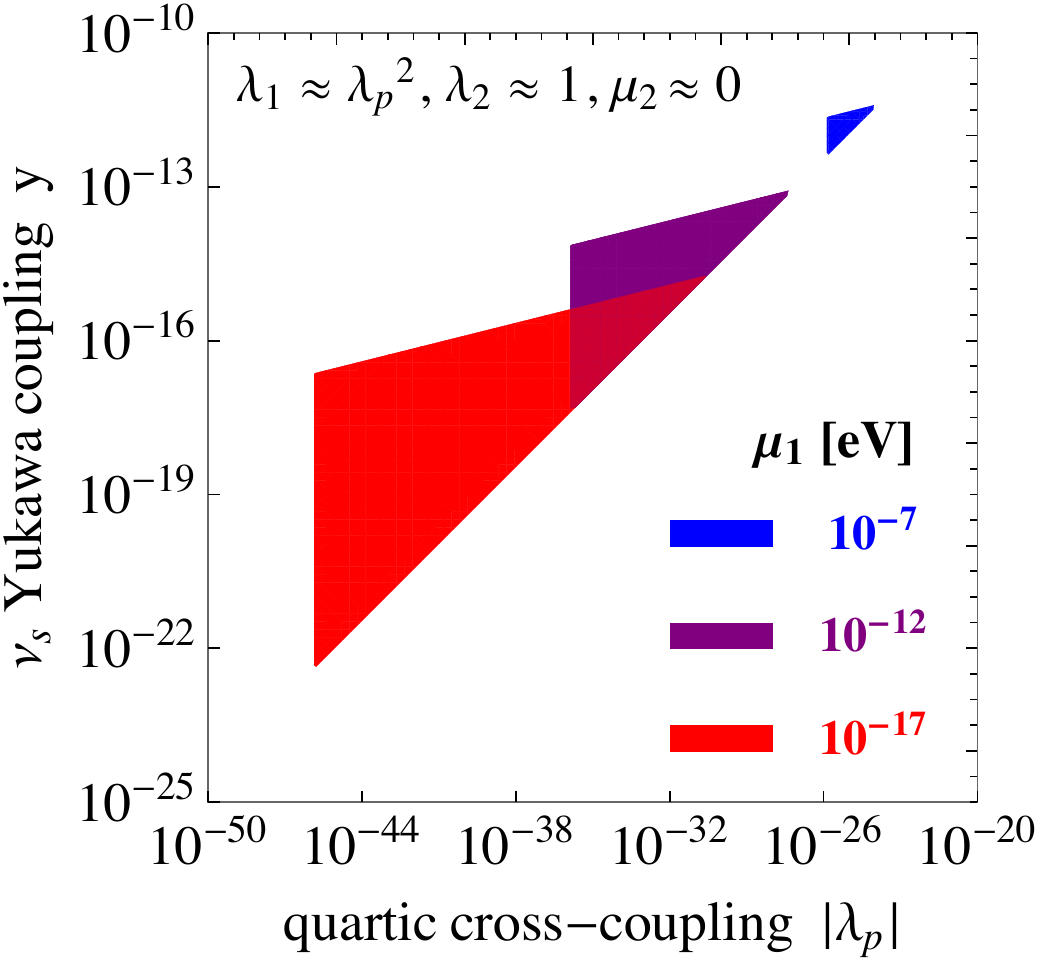}
  \caption{Parameter regions in which active and sterile neutrinos never
    recouple in the toy model given by \cref{eq:V-phi1-phi2,eq:L-ISB}.
    Different colors corresponds to different values of the scalar mass
    parameter $\mu_1$, as indicated in the legend.  We have taken
    $\lambda_1\simeq \lambda_p^2$, $\lambda_2 \simeq 1$, $\mu_2 \simeq 0$ to
    make this figure.}
  \label{fig:inverse-symmetry}
\end{figure}

The parameter space which satisfies all the above conditions is shown
in \cref{fig:inverse-symmetry} for various choices  of $\mu_1$.
We see that in this minimal toy model the parameters $\lambda_p$ and $y$
need to have rather extreme values. Nevertheless, the model serves
as a proof of principle that inverse symmetry breaking provides a
viable mechanism for preventing $\nu_s$ production in the early Universe.

To constrain the favored parameter regions of this model more quantitatively,
it would be necessary to compute the effective temperature-dependent potential
$V_\text{eff}$ and follow its evolution through cosmological history, for
instance using the public software package {\tt CosmoTransitions}
\cite{Wainwright:2011kj, Kozaczuk:2014kva}. For a given set of model
parameters, this would lead to a prediction for the temperature-dependent mass
of the sterile neutrinos, which could be plugged into the Boltzmann equations
governing their production to determine their final abundance. The relevant
contributions to $V_\text{eff}$ are, besides the tree level terms given in
\cref{eq:V-phi1-phi2}, the temperature-independent Coleman-Weinberg corrections
\cite{Coleman:1973jx, Quiros:1999jp}, the one-loop finite-temperature corrections
\cite{Dolan:1973qd}, and the resumed higher-order ``daisy'' terms
\cite{Carrington:1991hz}. As our goal here is merely to illustrate the
phenomenological viability of sterile neutrino models with inverse symmetry
breaking, this computation is far beyond the scope of the present work.

This model is an example for a more general class of models exhibiting inverse
symmetry breaking, where there is greater symmetry at lower temperatures as
opposed to the usual scenario where symmetries are restored at higher
temperatures~\cite{Weinberg:1974hy, Mohapatra:1979vr, Bajc:1999cn}. There are
other implementations of this mechanism, for instance Weinberg's $O(N_1)\times
O(N_2)$ scalar models~\cite{Weinberg:1974hy}, which break to $O(N_1-1)\times
O(N_2)$ at high temperature. At the non-perturbative level the symmetry may get
restored at very high temperatures and the parameter space available for such
inverse symmetry breaking is smaller than what is suggested by a 1-loop
perturbative treatment~\cite{Gavela:1998ux, Bimonte:1999tw, Pinto:1999pg}.
However, for our purposes it is sufficient that a phase of broken symmetry
exists at intermediate temperatures.

\subsection{Recoupling happens below MeV but CMB bounds on neutrino mass are avoided}
\label{sec:extra-dof}

An alternative way of reconciling sterile neutrinos with cosmology is to
tolerate their production at $T_\gamma < 1$~MeV, but to invoke extra
degrees of freedom to evade constraints on $\sum m_\nu$.

\emph{(i) Extra relativistic degrees of freedom to avoid structure formation
bounds.} At intermediate couplings (blue/orange vertically striped region in
\cref{fig:paramspace}), eV-scale sterile neutrinos with secret interactions are
constrained only by structure formation bounds on $\sum m_\nu$.  One way to
avoid these is to introduce several additional sterile states, also charged
under $U(1)_s$ and with not too small mixing with active neutrinos, but with
masses $\ll 1$\,eV. When secret interactions recouple active and sterile states
at temperatures $< \text{MeV}$, the energy density in the neutrino sector is
evenly distributed among all neutrino states. If the number of nearly massless
states is sufficiently large, only a small fraction of energy will remain for
the eV-scale state.  More precisely, by adding $n$ massless states in addition
to the three active neutrinos and the one eV-scale states, the energy density
after recoupling will be $3 \rho_\text{SM} / (4 + n)$ in each state, where
$\rho_\text{SM}$ is the energy density of each active neutrino flavor in the
SM.  Correspondingly, the effective bound on $\sum m_\nu$ is weakened by a
factor $3/(4+n)$. We see that, in order to reconcile a 1\,eV sterile neutrino
with the limit $\sum m_\nu < 0.23$\,eV \cite{Ade:2015xua}, we need to add $n
\geq 9$ massless states.

\emph{(ii) Extra relativistic degree of freedom for enhanced free-streaming.}
At large $\alpha_s$ (red region in \cref{fig:paramspace}, where the $\sum
m_\nu$ constraint is avoided because $\nu_s$ start to free-stream only very
late, the main problem faced by the vanilla model is that also at least one of
the active neutrinos will start to free-stream too late. Again, this problem
could be avoided by adding one extra relativistic species. This could be for
instance a second, nearly massless, sterile neutrino that partially thermalizes
before neutrino decoupling, or it could be the $A'$ boson itself, provided it
is nearly massless.  This scenario would predict $N_\text{eff}$ slightly larger
than the SM value, but possibly still consistent with constraints. On the other
hand, the extra free-streaming provided by the additional species is likely to
improve the fit to CMB and structure formation data nevertheless. A detailed
investigation of the viability of such a scenario requires a full fit to CMB
and large scale structure data, which is left for future work.

\emph{(iii) Fast sterile neutrino decay.} Cosmological constraints on the
secret interactions scenario could also be avoided if the eV-scale sterile
neutrino decays fast enough.  In particular, if $\nu_s$ decays to nearly
massless states before the onset of structure formation at $T \sim
1$~\text{eV}, large scale structure observations can hardly probe  the impact
of $\nu_s$ mass.  An appealing possibility is to introduce two additional
(nearly) massless particles: one pseudo-Goldstone boson $\phi$, and a second
sterile neutrino $\nu_s'$. For an interaction vertex of the form $y \phi
(\bar\nu_s \gamma_5 \nu_s')$ with $y \gtrsim 10^{-13}$, the lifetime corresponding
to the decay $\nu_s \to \nu_s'+ \phi$ is shorter than the time scale of
recombination, avoiding the CMB bounds on both $\sum m_\nu$ and $N_\text{eff}$.
In scenarios of this type, the strong $\nu_s$ self-interaction induced by the
coupling $\phi (\bar \nu_s \gamma_5 \nu_s)$ is in itself enough to suppress
$\nu_s$ production before BBN~\cite{Archidiacono:2014nda}.  In other words,
$\phi$ can take the place of the gauge boson $A'$ in mediating secret
interactions.  Alternative decay scenarios, such as three-body decays $\nu_s
\to 2\nu_s' + \bar\nu_s'$ (via an off-shell massive $A'$) or  $\nu_s \to \nu_s'
+ \gamma $ cannot generate sufficiently fast $\nu_s$ decays without violating
cosmological constraints~\cite{Forastieri:2017oma}.

\section{Summary}
\label{sec:conclusions}

To summarize, we have assessed the status of models featuring light sterile
neutrinos $\nu_s$ with ``secret'' self-interactions mediated by a new gauge
boson $A'$.  Such models had originally been introduced as a way of reconciling
light sterile neutrinos (as motivated by the short baseline oscillation anomalies)
with cosmological constraints. Indeed, the effective
temperature-dependent potential generated by secret interactions can
efficiently suppress active--sterile neutrino mixing in the early Universe down
to temperatures $\ll \text{MeV}$.  At that time, SM weak interactions have
frozen out and the neutrino sector is fully decoupled from the photon bath, so that
the number of relativistic species $N_\text{eff}$ cannot change any more.
However, efficient collisional production of $\nu_s$ (at the expense of active
neutrino $\nu_a$) will occur as soon as the mixing angle suppression is lifted.
Of particular importance in this context are $A'$-mediated scattering processes
which can be strongly enhanced by the $s$-channel $A'$ resonance (for $M \sim
T_\nu$), and by collinear enhancement in the forward direction (for $M \ll T_\nu$).
For $\nu_s$ masses around 1~eV and vacuum mixing angles of order 0.1 (as
motivated by the short-baseline oscillation anomalies), the resulting
population of $\nu_s$ is large enough to violate the cosmological constraint on
$\sum m_\nu$.  Thus, for $m_s = 1$~eV and $\theta_0 = 0.1$, all values of the
$A'$ mass $M$ and the corresponding fine structure constant $\alpha_s$ are
disfavored.  Our results confirm earlier findings from
ref.~\cite{Cherry:2016jol}.  A possible loophole to these
arguments exists at very large values of $\alpha_s$. Namely, if secret
interactions are so strong that $\nu_s$ cannot free-stream, measurements of
$\sum m_\nu$ are not sensitive.  However, it has been shown in
ref.~\cite{Forastieri:2017oma} that in this case also active neutrino
free-streaming is reduced, in conflict with CMB bounds.

In the second part of the paper we have discussed several new mechanisms for
reconciling eV-scale sterile neutrinos with cosmology.  We have outlined a toy
model in which sterile neutrinos have initially a very large mass generated by
the vacuum expectation value (vev) of a new scalar field, so that their
production is kinematically forbidden.  Only very late in cosmological history,
a phase transition reduces the scalar vev to zero and the $\nu_s$ mass to
$\mathcal{O}(\text{eV})$.  At that time, collisional production is no longer
possible, so the cosmological $\nu_s$ abundance remains negligible to this day.
We have shown that this scenario is viable, but requires rather extreme values
for some of its coupling constants.

We have in addition discussed scenarios with several new relativistic degrees
of freedom with masses $\ll \text{eV}$.  As far as cosmological bounds are
concerned, these degrees of freedom behave like active neutrinos.  They can
either serve to deplete the $\nu_s$ abundance by equilibrating with them, or
they can act as an extra free-streaming species, compensating for the reduced
free-streaming of active neutrinos in the regime of very large $\alpha_s$.

Finally, we have argued that bounds on $\sum m_\nu$ can also be avoided in
scenarios in which the sterile neutrino has a fast
decay mode to active neutrinos plus a light boson. Such models are of particular
interest in the context of the LSND and MiniBooNE anomalies \cite{Bai:2015ztj}.

\section*{Acknowledgements}

It is a pleasure to thank Alessandro Mirizzi for useful discussions. 
 X.C. is supported by the ‘New Frontiers’ program of the Austrian Academy of Sciences. 
BD is partially supported by the Dept.\ of Science and Technology of the
Govt.\ of India through a Ramanujam Fellowship and by the Max Planck-Gesellschaft
through a Max Planck-Partnergroup. JK and NS have been supported by the German
Research Foundation (DFG) under Grant  Grant Nos. KO 4820/11, FOR 2239,
EXC-1098 (PRISMA) and by the European Research Council (ERC) under the European
Union's Horizon 2020 research and innovation programme (grant agreement No.
637506, ``$\nu$Directions'').  JK would like to thank CERN for hospitality and
support during the final months of this project.

\begin{table*}[!t]
  \caption{Dominant processes and cross sections for production of the mostly
  sterile mass eigenstate $\nu_4$.}
  \label{tab:proc}
  \begin{center}
    \setlength\doublerulesep{.2pt} 
    \setlength{\extrarowheight}{10pt}  
    \begin{tabular}{r@{\quad}p{3.5cm}p{13cm}}
      \toprule[1pt]
      & process & relativistic cross section \\[0.2cm]
      \hline
      \emph{(i)} & $e^- + e^+ \to \bar \nu_1 + \nu_4$  &
      $\sigma_{ep\bar{1}4} = \sin^2\theta_m \,
          {\pi \alpha^2 \over 48 s_W^4 c_W^4} \,
          {\sqrt{s} \over  \sqrt{s - 4m_e^2}} \,
          {(13 - 20 c_W^2 + 8 c_W^4) s + (23 - 40 c_W^2 + 16 c_W^4) m_e^2   \over m_Z^4}$ \\
      \emph{(ii)} & $e^- + \nu_1 \to e^- + \nu_4$ &
      $\sigma_{e1e4} = \sin^2\theta_m
          {\pi \alpha^2 \over 24 s_W^4 c_W^4} \,
          {(s - m_e^2)^2 \over m_Z^4} \,
          {(31 - 44 c_W^2 + 16 c_W^4) s^2 - 2 (7 - 11 c_W^2 + 4 c_W^4) s \, m_e^2
                   + 4 (1 - c_W^2 )^2 m_e^4   \over   s^3}$ \\
      \emph{(iii)} & $e^+ + \nu_1 \to e^+ + \nu_4$ &
      $\sigma_{p1p4} = \sin^2\theta_m
          {\pi \alpha^2 \over 24 s_W^4 c_W^4} \,
          {(s - m_e^2)^2 \over m_Z^4} \,
          {(21 - 36 c_W^4 + 16 c_W^4) s^2 - (9 - 18 c_W^2 + 8 c_W^4) s \, m_e^2
                   + (3 - 2 c_W^2 )^2 m_e^4   \over   s^3}$ \\
      \emph{(iv)} & $\nu_1 + \nu_1 \to \nu_1 + \nu_4$ &
      $\sigma_{1114} = \sin^2\theta_m
          {\pi \alpha^2 \over s_W^4 c_W^4} \, {s \over 2 m_Z^4} $ \\  
      \emph{(v)} & $\bar\nu_1 + \nu_1 \to \bar\nu_1 + \nu_4$ &
      $\sigma_{\overline{1}1\overline{1}4} = \sin^2\theta_m
          {\pi \alpha^2 \over s_W^4 c_W^4} \, {s \over 3 m_Z^4}$ \\[.2cm]
      \hline 
      \emph{(vi)} & $\bar\nu_4 +\nu_1 \to \bar\nu_4 + \nu_4$ &
      $\sigma_{\overline{4}1\overline{4}4} = \sin^2\theta_m
          {4\pi \alpha_s^2 \over 3} \,
          {(3 s^2 + 10 M^2 s - 12 M^4 )s + 12 M^2 (s^2 - M^4) \log(1 + s / M^2) \over
          M^2s [(s - M^2)^2 + M^2 \Gamma_M^2]}$ \\
      \emph{(vii)} & $\nu_4 + \nu_1 \to \nu_4 + \nu_4$ &
      $\sigma_{4144} = \sin^2\theta_m
          {4\pi \alpha_s^2} \,
          {(s + 2 M^2) s + 2 M^2 (s + M^2) \log(1 + s/M^2) \over
          M^2 (s + M^2) (s + 2 M^2)}$  \\[.2cm]
      \bottomrule[1pt]
    \end{tabular}
  \end{center}
\end{table*}

\appendix
\section{Production Rate of Sterile Neutrinos}
\label{sec:appendix}

As discussed in \cref{sec:constraints}, we have taken into account seven different
processes in the computation of the sterile neutrino recoupling temperature.
Here, in Table~\ref{tab:proc}, we list the corresponding cross sections in the $1+1$ flavor approximation
and to leading order in the mixing angle.
Of course, the corresponding CP-conjugate processes contribute with the same
cross sections.  Note that process \emph{(iv)} has identical initial state
particles, a fact that needs to be properly taken into account when computing
the rate for this process by integrating over the distribution of initial state
$\nu_e$.  In process \emph{(vi)}, which can be mediated by an $s$-channel $A'$,
we need to take into account the non-zero width of $A'$, which is given by
$\Gamma_M = \alpha_s M/ 3$ for Dirac $\nu_s$.

The production rate of sterile neutrinos, normalized to the volume element
occupied by an active neutrino as explained in \cref{sec:Trec}, is given by 
\begin{align}
  \Gamma_s = \frac{c_{QZ}}{n^\text{eq}_{\nu_a}}
               \sum_i \int \! d^3p_1 \, d^3p_2 \,
                           f_i(\vec p_1) f_i(\vec p_2)
                           \times \sigma_i(s) v_\text{M\o l} \,,
  \label{eq:Gamma-s-app-1}
\end{align}
where the sum runs over the seven production processes listed above, the
integral is over the 3-momenta of the initial state particles, and the
prefactor $c_{QZ}$ accounts for quantum Zeno suppression (see \cref{sec:Trec}
for details).  The M\o ller velocity $v_\text{M\o l}$ reduces to the relative
velocity of the two colliding particles in the non-relativistic limit (see
ref.~\cite{Gondolo:1990dk} for details).  The momentum distribution functions
of the initial state particles $f_i(\vec p_{1,2})$ have a Fermi-Dirac
shape as we only consider fermionic processes. The number density of active
neutrinos $n^\text{eq}_{\nu_a}$ is the integral over the corresponding Fermi-Dirac
distribution for one massless degree of freedom. Abbreviating the integral
by introducing the notation $\langle \cdot \rangle$ for the momentum-averaged
cross section, we obtain \cref{eq:Gamma-s}.

The characteristic temperatures of the initial state particles are
$T_\gamma$ for electrons and positrons, $T_\nu$ for active neutrinos, and
$T_s$ for sterile neutrinos. As an approximation, we assume the same relation
between $T_\gamma$ and $T_\nu$ as in the SM.  The only exception is that for
processes \emph{(ii)} and \emph{(iii)}, which involve both a charged lepton and
a neutrino in the initial state, we set $T_\nu = T_\gamma$ despite the
temperature difference between them after $e^+e^-$ annihilation.
This approximation, which makes the
numerical evaluation of $\Gamma_s$ easier, is justified by the fact
that after BBN electrons quickly become decoupled, so that processes mediated
by SM $W$ and $Z$ bosons should not play an important role any longer. 
In the case of process \emph{(iv)}, which is initiated
by two identical particles, care must be taken to restrict the integration domain
such that double-counting of initial state phase space is avoided. (Or, alternatively,
an extra factor of $1/2$ needs to be included in the integrand.)

\begin{widetext}
The integrals in \cref{eq:Gamma-s-app-1} can be partially evaluated
\cite{Gondolo:1990dk}. With the definitions $E_\pm = E_1 \pm E_2$, the result
is
\begin{align}
  \Gamma_s = \frac{c_{QZ}}{n^\text{eq}_{\nu_a}} \,
             2  \pi^2 \, \sum_i \int \! ds \, dE_+ \, dE_-
               f_i\bigg({E_+ + E_- \over 2}\bigg)
               f_i\bigg({E_+ - E_- \over 2}\bigg)
               \times \sigma_{i}(s) \, \sqrt{[s - (m_1 + m_2)^2]
                                            [s - (m_1 - m_2)^2] \over s} \,,
\end{align}
with the integration boundaries
\begin{gather}
  s \geq (m_1+m_2)^2 \,, \hspace{2cm}
  E_+ \geq \sqrt{s} \,, \\[0.2cm]
  |E_- - E_+ \frac{m_1^2-m_2^2}{s}| \leq
    {\sqrt{(E_+^2-s) [s-(m_1 + m_2)^2][s-(m_1 - m_2)^2]} \over s} \,.
\end{gather}
We use the condition
\begin{align}
  \Gamma_s(T_\text{rec})  =  H(T_\text{rec}) \,
  \label{eq:Gamma-s-cond}
\end{align}
to numerically decide whether and at what recoupling temperature, $T_\text{rec}$,  the sterile $\nu_s$ can be brought into the thermal equilibrium with active neutrinos.
\end{widetext}

\bibliographystyle{JHEP}
\bibliography{refs}

\end{document}